\documentclass[a4paper, 11pt]{article}
\pdfoutput=1
\usepackage{jheppub}

\usepackage{verbatim} 
\newcommand{\Cbb}{\ensuremath{\mathbb C} }
\newcommand{\Ibb}{\ensuremath{\mathbb I} }
\newcommand{\cA}{\ensuremath{\mathcal A} }
\newcommand{\cD}{\ensuremath{\mathcal D} }
\newcommand{\cDbar}{\ensuremath{\overline{\mathcal D}} }
\newcommand{\cF}{\ensuremath{\mathcal F} }
\newcommand{\cFbar}{\ensuremath{\overline{\mathcal F}} }
\newcommand{\cN}{\ensuremath{\mathcal N} }
\newcommand{\cO}{\ensuremath{\mathcal O} }
\newcommand{\cP}{\ensuremath{\mathcal P} }
\newcommand{\cQ}{\ensuremath{\mathcal Q} }
\newcommand{\cU}{\ensuremath{\mathcal U} }
\newcommand{\cUbar}{\ensuremath{\overline{\mathcal U}} }
\newcommand{\al}{\ensuremath{\alpha} }
\newcommand{\be}{\ensuremath{\beta} }
\newcommand{\de}{\ensuremath{\delta} }
\newcommand{\eps}{\ensuremath{\epsilon} }

\newcommand{\lalat}{\ensuremath{\lambda_{\rm lat}} }
\newcommand{\ka}{\ensuremath{\kappa} }
\newcommand{\muhat}{\ensuremath{\widehat \mu} }
\newcommand{\etabar}{\ensuremath{\overline{\eta}} }
\newcommand{\phibar}{\ensuremath{\overline{\phi}} }
\newcommand{\psibar}{\ensuremath{\overline{\psi}} }
\newcommand{\glN}{\ensuremath{\mathfrak{gl}(N, \Cbb)} }
\newcommand{\slN}{\ensuremath{\mathfrak{sl}(N, \Cbb)} }
\newcommand{\UN}{\ensuremath{\mbox{U(}N\mbox{)}} }
\newcommand{\SUN}{\ensuremath{\mbox{SU(}N\mbox{)}} }
\newcommand{\Uone}{\ensuremath{\mbox{U(1)}} }
\renewcommand{\Re}{\ensuremath{\mbox{Re}\;} }
\renewcommand{\Im}{\ensuremath{\mbox{Im}\;} }
\newcommand{\lsim}{\ensuremath{\lesssim} }
\newcommand{\X}{\ensuremath{\!\times\!} }
\newcommand{\nn}{\nonumber}
\newcommand{\pf}{\ensuremath{\mbox{pf}\,} }
\newcommand{\Tr}[1]{\ensuremath{\mbox{Tr}\left[ #1 \right]} }
\newcommand{\vev}[1]{\ensuremath{\left\langle #1 \right\rangle} }
\newcommand{\deriv}[2]{\ensuremath{\frac{d #1}{d #2}} }
\newcommand{\eq}[1]{eq.~\ref{#1}}
\newcommand{\fig}[1]{figure~\ref{#1}}
\newcommand{\tab}[1]{table~\ref{#1}}
\newcommand{\secref}[1]{section~\ref{#1}}
\newcommand{\refcite}[1]{ref.~\cite{#1}}
\def\figheight{7 cm}

\title{Lifting flat directions in lattice supersymmetry}

\author{Simon~Catterall}
\author{and David~Schaich}
\affiliation{Department of Physics, Syracuse University, Syracuse, New York 13244, United States}
\emailAdd{smcatter@syr.edu}
\emailAdd{dschaich@syr.edu}

\abstract{ 
  We present a procedure to improve the lattice definition of $\cN = 4$ supersymmetric Yang--Mills theory.
  The lattice construction necessarily involves U(1) flat directions, and we show how these can be lifted without violating the exact lattice supersymmetry.
  The basic idea is to modify the equations of motion of an auxiliary field, which determine the moduli space of the system.
  Applied to numerical calculations, the resulting improved lattice action leads to dramatically reduced violations of supersymmetric Ward identities and much more rapid approach to the continuum limit.
}
\arxivnumber{1505.03135}

\begin{document}
\maketitle
\flushbottom

\section{\label{sec:intro}Introduction} 
In recent years there has been tremendous progress in constructing supersymmetric lattice gauge theories, as reviewed by \refcite{Catterall:2009it} and references therein.
In particular, a lattice formulation of four-dimensional $\cN = 4$ supersymmetric Yang--Mills (SYM) theory has been constructed, which exactly preserves a single twisted supercharge \cQ at non-zero lattice spacing~\cite{Kaplan:2005ta, Unsal:2006qp, Catterall:2007kn, Damgaard:2008pa, Catterall:2011pd, Catterall:2012yq, Catterall:2013roa, Catterall:2014vka, Catterall:2014mha, Schaich:2014pda, Catterall:2014vga}.\footnote{Other approaches to studying $\cN = 4$ SYM numerically include refs.~\cite{Ishii:2008ib, Ishiki:2008te, Ishiki:2009sg, Hanada:2010kt, Honda:2011qk, Honda:2013nfa, Hanada:2013rga}.}
Initial studies, both analytical and numerical, have shown that this lattice theory possesses several remarkable properties:
\begin{itemize}
  \item The renormalization of the theory is highly constrained by gauge invariance, the exact \cQ supersymmetry and a global $S_5$ symmetry.  These symmetries ensure that at most a single marginal coupling may need to be tuned to correctly recover all 16 supercharges and the SU(4) R~symmetry of $\cN = 4$ SYM in the continuum limit~\cite{Catterall:2013roa, Catterall:2014mha}.
  \item The moduli space of the lattice theory is the same as that of the continuum theory, and is not lifted to any order of lattice perturbation theory~\cite{Catterall:2011pd}.
  \item The \be function vanishes at one loop in lattice perturbation theory~\cite{Catterall:2011pd}.
  \item Although the pfaffian that results from integrating over the fermion fields in the lattice path integral is potentially complex, pfaffian-phase-quenched computations display no indication of a sign problem~\cite{Catterall:2014vka}.  This allows numerical computations to employ conventional Monte Carlo algorithms~\cite{Schaich:2014pda}.
  \item The static potential of the lattice theory is coulombic at both weak and strong coupling, in agreement with continuum expectations~\cite{Catterall:2012yq, Catterall:2014vka, Catterall:2014vga}.
\end{itemize}

Given the central role $\cN = 4$ SYM plays in the AdS/CFT correspondence that relates it to quantum gravity, it is important to carry out large-scale lattice calculations to investigate the theory away from the regime of weak coupling and for arbitrary numbers of colors $N$.
Despite the encouraging results described above, the formal action constructed in refs.~\cite{Kaplan:2005ta, Catterall:2007kn} possesses certain features that obstruct such numerical calculations.
The primary issue, explored in \refcite{Catterall:2014vka}, is that this lattice system possesses flat directions that destabilize the vacuum of the theory.
In particular, the theory necessarily involves $\UN = \SUN \otimes \Uone$ gauge invariance, and flat directions associated with the U(1) sector produce especially severe lattice artifacts.
This behavior may be understood by noting that constant U(1) shifts of the fields $X(x) \to X(x) + c\Ibb$ leave the action $S$ unchanged for {\em any} $x$-independent field configurations $X(x)$ even when $S[X] \neq 0$.
This invariance should be contrasted with that of the usual SU($N$) flat directions that correspond to constant shifts by elements of the Cartan subalgebra on $x$-independent field configurations for which the action $S = 0$.
In other words, the SU($N$) flat directions involve only supersymmetric vacuum solutions while the U(1) flat directions can play a more general role.
Of course the continuum theory admits a trivial truncation to SU($N$), but this is not true for the lattice theory at non-zero lattice spacing.

In \refcite{Catterall:2014vka} the fluctuations of these U(1) modes were controlled in a way that violated the exact supersymmetry.
Although the resulting soft supersymmetry breaking was under control for 't~Hooft couplings $\lalat \lsim 5$, and suppressed $\propto 1 / N^2$, it limited the exploration of stronger couplings and produced a slow approach to the continuum limit~\cite{Catterall:2014vga}.
In this paper we show how these problems may be solved through a construction that regulates the U(1) modes in a manner compatible with the \cQ supersymmetry.
The basic idea is to modify the equations of motion of an auxiliary field, which determine the moduli space of the system.
By maintaining the \cQ supersymmetry we ensure that $\cQ$-invariant observables such as the partition function can still be computed by restricting the functional integrals to a moduli space that has been modified to include only the SU($N$) sector.

We begin in the next section with a brief review of the difficulties faced by the formal supersymmetric lattice action, summarizing how these were addressed by the unimproved action used in the past, at the cost of soft supersymmetry breaking.
In \secref{sec:deform} we describe the general framework of $\cQ$-invariant modifications to the moduli equations, which we use to construct our improved lattice action.
We test this improved action in \secref{sec:tests}, finding that it dramatically reduces violations of Ward identities on small $L^4$ lattice volumes and exhibits much more rapid approach to the continuum limit as $L$ increases.
Although the improved action is almost twice as computationally expensive as the unimproved action, these benefits are so substantial that we recommend using the improved action in future studies of lattice $\cN = 4$ SYM.

One may also consider alternate modifications of the moduli equations, for which this procedure produces other possible lattice actions.
We discuss one interesting example in appendix~A, which turns out not to behave as well as the improved action we focus on.
Although further explorations in this direction may be worthwhile, the improved action we present here appears hard to beat in the context of practical numerical calculations.
In particular, in appendix~B we argue that the symmetries of the lattice system forbid all dimension-5 operators, which allows the new action to be effectively $\cO(a)$ improved.
Finally, in appendix~C we exploit the improved action to provide a brief update on the absence of a sign problem in lattice $\cN = 4$ SYM.

In the course of this paper we will discuss several different lattice actions, which we briefly define here to avoid confusion:
\begin{itemize}
  \setlength{\itemsep}{1 pt}
  \setlength{\parskip}{0 pt}
  \setlength{\parsep}{0 pt}
  \item The {\bf formal} action $S_{\rm formal}$ directly discretizes continuum twisted $\cN = 4$ SYM, exactly preserving the \cQ supersymmetry but suffering from difficulties summarized in the next section.
  \item The {\bf unimproved} action $S_{\rm unimp}$ used in refs.~\cite{Catterall:2014vka, Schaich:2014pda, Catterall:2014vga} stabilizes numerical calculations through two new terms, each of which softly breaks supersymmetry.
  \item The {\bf improved} action $S_{\rm imp}$ introduced in this work replaces one of the two soft $\cQ$-breaking terms of the unimproved action by a supersymmetric construction, removing the dominant source of supersymmetry breaking.
  \item The {\bf over-constrained} action $S_{\rm over}$ discussed in appendix~A implements all necessary stabilizations in a $\cQ$-invariant manner, but this turns out to introduce new difficulties.
\end{itemize}

\section{\label{sec:stab}Difficulties of the formal supersymmetric lattice action} 
The starting point for lattice $\cN = 4$ SYM is the direct and exactly $\cQ$-supersymmetric discretization of the continuum twisted action,
\begin{align}
  S_{\rm formal}  & = S_{\rm exact} + S_{\rm closed}                                                                                                                     \label{eq:S0}      \\
  S_{\rm exact}  & = \frac{N}{2\lalat} \sum_n \Tr{\cQ \left(\chi_{ab}(n)\cD_a^{(+)}\cU_b(n) + \eta(n) \cDbar_a^{(-)}\cU_a(n) - \frac{1}{2}\eta(n) d(n) \right)} \label{eq:Sexact0} \\
  S_{\rm closed} & = -\frac{N}{8\lalat} \sum_n \Tr{\eps_{abcde}\ \chi_{de}(n + \muhat_a + \muhat_b + \muhat_c) \cDbar_c^{(-)} \chi_{ab}(n)},                    \label{eq:Sclosed}
\end{align}
with repeated indices summed.
All fields $X = \sum_{A = 1}^{N^2} T^A X^A$ are expanded on a basis of anti-hermitian generators of the group, $\Tr{T^A T^B} = -\de^{AB}$.
Here $\cU_a$ with $a = 0 \cdots 4$ are complexified gauge links that also contain the scalar fields and $d$ is a bosonic auxiliary field, while $\eta$ and $\chi_{ab} = -\chi_{ba}$ account for 11 of the 16 twisted fermion field components.
The lattice finite-difference operators are~\cite{Catterall:2007kn, Damgaard:2008pa}
\begin{align}
  \cD_a^{(+)} f_b(n)       & = \cU_a(n) f_b(n + \muhat_a) - f_b(n) \cU_a(n + \muhat_b)                        \cr
  \cDbar_a^{(-)} f_a(n)    & = f_a(n) \cUbar_a(n) - \cUbar_a(n - \muhat_a) f_a(n - \muhat_a)   \label{eq:diff} \\
  \cDbar_a^{(+)} f_b(n)    & = \cUbar_a(n + \muhat_b) f_b(n) - f_b(n + \muhat_a) \cUbar_a(n)                  \cr
  \cDbar_c^{(-)} f_{ab}(n) & = f_{ab}(n + \muhat_c) \cUbar_c(n) - \cUbar_c(n + \muhat_a + \muhat_b)f_{ab}(n). \nn
\end{align}

This lattice action is invariant under the \cQ supersymmetry
\begin{align}
  & \cQ\; \cU_a(n) = \psi_a(n)            & & \cQ\; \psi_a(n) = 0                   \cr
  & \cQ\; \chi_{ab}(n) = -\cFbar_{ab}(n)  & & \cQ\; \cUbar_a(n) = 0 \label{eq:susy} \\
  & \cQ\; \eta(n) = d(n)                  & & \cQ\; d(n) = 0                        \nn
\end{align}
where $\psi_a$ are the remaining 5 fermion field components and the complexified field strength is $\cFbar_{ab} = \cDbar_a^{(+)} \cUbar_b = -\cFbar_{ba}$.
From \eq{eq:susy} we see $\cQ^2 = 0$, which establishes the invariance of $S_{\rm exact}$.
The invariance of $S_{\rm closed}$ follows from an exact lattice Bianchi identity.
If we integrate out the bosonic auxiliary field $d$ through its equation of motion $d = \cDbar_a^{(-)}\cU_a$, \eq{eq:Sexact0} becomes
\begin{align}
  S_{\rm exact} & = \frac{N}{2\lalat} \sum_n \mbox{Tr}\bigg[-\cFbar_{ab}(n)\cF_{ab}(n) - \chi_{ab}(n) \cD_{[a}^{(+)}\psi_{b]}^{\ }(n) - \eta(n) \cDbar_a^{(-)}\psi_a(n) \label{eq:Sexact} \\
            & \hspace{10 cm} + \frac{1}{2}\left(\cDbar_a^{(-)}\cU_a(n)\right)^2\bigg].                                                                                                   \nn
\end{align}

This $\cQ$-invariant action possesses certain features that interfere with numerical calculations.
These features are direct consequences of the structure of the continuum theory and the requirement of exact \cQ supersymmetry at non-zero lattice spacing.
First note that \eq{eq:susy} forces the complexified gauge link variables $\cU_a$ to reside in the algebra \glN of the group U($N$), in correspondence with $\psi_a$.\footnote{The lattice path integral over gauge fields that live in the algebra involves the flat measure rather than the usual Haar measure familiar from lattice QCD.  This formulation remains gauge invariant since the Jacobian resulting from a gauge transformation on $D\cU$ cancels against the corresponding quantity for $D\cUbar$.}
Therefore targeting the correct continuum theory requires that the links have the expansion $\cU_b(n) = \frac{1}{a}\Ibb_N + \cA_b(x) + \cO(a)$ in some suitable gauge, where $a$ is the lattice spacing and $\cA_b$ is the complexified gauge field of the twisted continuum theory.
We can interpret the unit matrix in this expansion as the vacuum expectation value (vev) of an imaginary U(1) mode, implying that we must lift the corresponding flat directions.
To do so we need to add to the action a scalar potential, typically $\sum_a \left(\frac{1}{N} \Tr{\cU_a \cUbar_a} - 1\right)^2$, which will also regulate the usual SU($N$) flat directions.

However, such a scalar potential does not affect real U(1) modes, which cancel out of $\cU_a \cUbar_a$.
While these U(1) modes simply decouple in the continuum theory, in \refcite{Catterall:2014vka} we showed that on the lattice they lead to unphysical confinement at strong coupling, indicated by the condensation of lattice-artifact U(1) monopoles.
We therefore need to add another regulator to the action, which we take to be some function of $\left(\det \cP_{ab} - 1\right)$, where
\begin{equation}
  \label{eq:plaquette}
  \cP_{ab}(n) = \cU_b(n) \cU_a(n + \muhat_b) \cUbar_b(n + \muhat_a) \cUbar_a(n)
\end{equation}
is the oriented plaquette in the $a$--$b$ plane.
Since this term does not affect the SU($N$) flat directions, we still need to retain the scalar potential to fully stabilize numerical calculations.

Finally, the fermion action possesses an exact U(1) zero mode corresponding to the shift symmetry $\eta \to \eta + c\Ibb_N$, with $c$ a constant Grassmann parameter.
We typically lift this zero mode by imposing anti-periodic (thermal) temporal boundary conditions (BCs) for the fermion fields, which explicitly breaks the \cQ supersymmetry since the bosonic fields are always periodic in all directions.
This source of supersymmetry breaking is controlled by the temporal extent of the lattice, $N_t$.
Previous studies observed negligible effects even for $N_t = 4$~\cite{Catterall:2014vka}, though this may be a non-universal feature of the unimproved action then employed.
We will revisit this issue for the improved action in \secref{sec:tests}.

In summary the formal lattice theory suffers from problems that have their origin in a U(1) sector that necessarily arises in the lattice construction and that does not decouple from the physical SU($N$) sector at non-zero lattice spacing.
This U(1) sector possesses flat directions that can destabilize lattice calculations.
To avoid instabilities and ensure that the lattice theory possesses the correct continuum limit we must regulate these U(1) flat directions.
In the past, we did so by adding the simplest possible scalar potential and plaquette determinant terms directly to \eq{eq:S0}, as
\begin{align}
  S_{\rm unimp} & = S_{\rm formal} + S_{\rm soft}                                                                                                                 \label{eq:Ssoft} \\
  S_{\rm soft}  & = \frac{N}{2\lalat} \mu^2 \sum_n \sum_a \left(\frac{1}{N} \Tr{\cU_a(n) \cUbar_a(n)} - 1\right)^2 + \ka \sum_n \sum_{a < b} \left|\det \cP_{ab}(n) - 1\right|^2. \nn
\end{align}
Since these terms involve only bosonic and not fermionic fields, they both introduce soft supersymmetry breaking that vanishes in a controlled way in the limit $(\mu, \ka) \to (0, 0)$~\cite{Catterall:2014vka, Catterall:2014vga}.
We now describe a new method that allows us to introduce such deformations in a manner compatible with the exact \cQ supersymmetry.

\section{\label{sec:deform}Supersymmetric deformations of the moduli space} 
\subsection{General framework} 
The key observation reported by this paper is that one can modify the structure of the moduli equations to enforce new vacuum conditions without breaking the \cQ supersymmetry.
This procedure can be used to regulate the problematic U(1) flat directions.
We proceed by deforming the $\cQ\; \eta \cDbar^{(-)}_a \cU_a$ term in $S_{\rm exact}$ (\eq{eq:Sexact0}),
\begin{equation}
  \label{eq:generic_deform}
  \cQ\; \Tr{\eta(n) \left(\cDbar_a^{(-)}\cU_a(n)\right)} \to \cQ\; \Tr{\eta(n) \left(\cDbar_a^{(-)}\cU_a(n) + G\cO(n)\Ibb_N\right)}
\end{equation}
where $G$ is a new tunable coupling and $\cO(n)$ is a gauge-invariant bosonic operator.
Although we only recover lattice $\cN = 4$ SYM in the limit $G \to 0$, the deformed $S_{\rm exact}^{\prime}$ part of the lattice action remains \cQ exact.

Performing the \cQ variation in \eq{eq:generic_deform} produces
\begin{equation}
  \begin{split}
    \cQ\; \Tr{\eta(n) \left(\cDbar_a^{(-)}\cU_a(n) + G\cO(n)\Ibb_N\right)} =\ & \Tr{d(n) \left(\cDbar_a^{(-)}\cU_a(n) + G\cO(n)\Ibb_N\right)} \\
    & - \Tr{\eta(n) \left(\cDbar_a^{(-)}\psi_a(n) + G \cQ\; \cO(n) \Ibb_N\right)}.
  \end{split}
\end{equation}
In combination with the $\cQ\; \eta d = d^2$ term in \eq{eq:Sexact0} we obtain the modified equations of motion
\begin{equation}
  \label{eq:EOM}
  d(n) = \cDbar_a^{(-)}\cU_a(n) + G \cO(n)\Ibb_N.
\end{equation}
Since $\cQ\; \eta = d$, this deformation also modifies a \cQ Ward identity,
\begin{equation}
  \label{eq:Ward}
  \vev{\sum_n \Tr{\cQ\; \eta(n)}} = \vev{\sum_n \Tr{\cDbar_a \cU_a(n)} + NG \sum_n \cO(n)} = NG \vev{\sum_n \cO(n)} = 0,
\end{equation}
where the first term vanishes due to the sum over the lattice volume and the structure of the finite-difference operator, \eq{eq:diff}.
After integrating out the auxiliary field, only the second line of \eq{eq:Sexact} for $S_{\rm exact}$ changes,
\begin{align}
  S_{\rm exact}^{\prime} & = \frac{N}{2\lalat} \sum_n \mbox{Tr}\bigg[-\cFbar_{ab}(n)\cF_{ab}(n) - \chi_{ab}(n) \cD_{[a}^{(+)}\psi_{b]}^{\ }(n) - \eta(n) \cDbar_a^{(-)}\psi_a(n) \label{eq:Sexact_deform} \\
                         & \hspace{5 cm} + \frac{1}{2}\left(\cDbar_a^{(-)}\cU_a(n) + G \cO(n)\Ibb_N\right)^2 - G \eta(n) \cQ\; \cO(n)\bigg].                                                             \nn
\end{align}

\subsection{\label{sec:improved}Application to construct the improved action} 
Since the plaquette determinant term in $S_{\rm soft}$ (\eq{eq:Ssoft}) produces much larger soft supersymmetry breaking than does the scalar potential term~\cite{Catterall:2014vka}, we consider
\begin{equation}
  \label{eq:linear}
  \cO(n) = \sum_{a \neq b} \left(\det\cP_{ab}(n) - 1\right) = 2\Re\sum_{a < b}\left(\det\cP_{ab}(n) - 1\right),
\end{equation}
where the second equality follows from $\cP_{ba} = \cP_{ab}^*$.
With this $\cO(n)$, the Ward identity in \eq{eq:Ward} produces the single constraint
\begin{equation}
  \label{eq:detWard}
  \sum_n \sum_{a \neq b} \vev{\det\cP_{ab}(n) - 1} = 0.
\end{equation}
Invariance under discrete lattice translations and rotations then implies $\vev{\Re\det\cP_{ab}(n)} = 1$.
Although the deformation of the lattice moduli space does not constrain $\vev{\Im\det\cP_{ab}(n)}$, the same scalar potential in \eq{eq:Ssoft} that regulates the SU($N$) flat directions will also address this issue, by requiring $\frac{1}{N} \Tr{\cU_a(n) \cUbar_a(n)} \approx 1$.

To write down the improved lattice action we need the \cQ variation of the plaquette determinant, which follows from Jacobi's formula
\begin{equation}
  \deriv{\det\cU_a}{\al} = \left[\det\cU_a\right] \Tr{\cU_a^{-1} \deriv{\cU_a}{\al}}.
\end{equation}
Applying this to the determinant of \eq{eq:plaquette}, we obtain
\begin{align}
  \cQ \left[\det\cP_{ab}(n) - 1\right] = \left[\det\cP_{ab}(n)\right] \Tr{\cU_b^{-1}(n) \psi_b(n) + \cU_a^{-1}(n + \muhat_b) \psi_a(n + \muhat_b)}
\end{align}
so that the full improved action becomes
\begin{align}
  S_{\rm imp}            & = S_{\rm exact}^{\prime} + S_{\rm closed} + S_{\rm soft}^{\prime}                                                                                       \label{eq:Simp} \\
  S_{\rm exact}^{\prime} & = \frac{N}{2\lalat} \sum_n \mbox{Tr}\Bigg[-\cFbar_{ab}(n)\cF_{ab}(n) - \chi_{ab}(n) \cD_{[a}^{(+)}\psi_{b]}^{\ }(n) - \eta(n) \cDbar_a^{(-)}\psi_a(n)               \nn \\
                     & \hspace{4 cm} + \frac{1}{2}\left(\cDbar_a^{(-)}\cU_a(n) + G \sum_{a \neq b} \left(\det\cP_{ab}(n) - 1\right)\Ibb_N\right)^2\Bigg] - S_{\rm det}                         \nn \\
  S_{\rm det}            & = \frac{N}{2\lalat} G \sum_n \Tr{\eta(n)} \sum_{a \neq b} \left[\det\cP_{ab}(n)\right] \Tr{\cU_b^{-1}(n) \psi_b(n) + \cU_a^{-1}(n + \muhat_b) \psi_a(n + \muhat_b)} \nn \\
  S_{\rm soft}^{\prime}  & = \frac{N}{2\lalat} \mu^2 \sum_n \sum_a \left(\frac{1}{N} \Tr{\cU_a(n) \cUbar_a(n)} - 1\right)^2                                                                    \nn
\end{align}
with $S_{\rm closed}$ still given by \eq{eq:Sclosed}.
Written in this form, the \cQ invariance of $S_{\rm exact}^{\prime}$ is reflected by the presence of $\det\cP$ in both the bosonic action and the fermionic action.
The plaquette determinant terms in the fermionic action, written as $S_{\rm det}$ above, break the $\eta \to \eta + c\Ibb_N$ shift symmetry discussed in the previous section, and lift the U(1) fermion zero mode.
$S_{\rm det}$ involves the $\eta\psi_a$ fermion bilinear and can be interpreted as providing an effective fermion mass term, with the $\Tr{\eta}$ factor picking out the U(1) sector.

When implemented in the publicly available parallel software described in \refcite{Schaich:2014pda}, $S_{\rm det}$ also adds 40 inter-node data transfers to the fermion operator, bringing the total to more than 100 (many more than lattice QCD with Wilson or staggered fermions).
Although we no longer need to use anti-periodic temporal BCs to lift the U(1) fermion zero mode, we find that these BCs prevent the appearance of a sign problem that results if we use fully periodic BCs.
Mysteriously, we also measure most observables to be insensitive to the BCs and to the presence or absence of a sign problem.
We discuss these issues further in appendix~C and hope to gain further insight from future investigations.
As a practical matter, the anti-periodic temporal BCs significantly reduce the condition number of the fermion operator, accelerating numerical computations.
Even so, the additional terms and inter-node data transfers cause the improved action to be almost twice as computationally expensive as the unimproved action.

The linear dependence of the new supersymmetric deformation on $\left(\det\cP_{ab}(n) - 1\right)$ is another significant difference compared to the $\left|\det\cP_{ab}(n) - 1\right|^2$ used in the unimproved action, \eq{eq:Ssoft}.
While the latter term was designed to approximately project every plaquette to \slN at each site of the lattice, the new deformation focuses only on regulating the U(1) flat directions.
It was not obvious a priori that the linear form would suffice, but this does turn out to be the case.
However, the linear $\left(\det\cP - 1\right)$ deformation prevents us from also including the scalar potential in \eq{eq:linear} for $\cO(n)$.
When we do so we find that the scalar potential and plaquette determinant largely cancel each other out and fail to regulate the flat directions.

This leads us to retain the same soft supersymmetry breaking scalar potential as used in the past.
Although this $S_{\rm soft}^{\prime}$ does much less violence to \cQ than did the old plaquette determinant term, it is enticing to consider the possibility of obtaining a fully $\cQ$-invariant stabilized lattice action.
In order to do so we must make both the scalar potential and plaquette determinant deformations non-negative at each lattice site, for example by switching back to the $\left|\det\cP_{ab} - 1\right|^2$ form used in \eq{eq:Ssoft}.
We present such a construction in appendix~A, finding that the resulting action encounters new difficulties and does not behave as well as the improved action in \eq{eq:Simp}.

\section{\label{sec:tests}Tests of the improved action} 
\subsection{\label{sec:phase}Lattice phase diagrams} 
In this section we present numerical evidence that the improved action does in fact produce dramatic improvements compared to the $S_{\rm unimp}$ used in refs.~\cite{Catterall:2014vka, Schaich:2014pda, Catterall:2014vga}.
Each action involves two auxiliary couplings: $(\mu, G)$ for $S_{\rm imp}$ and $(\mu, \ka)$ for $S_{\rm unimp}$.
Since we only recover $\cN = 4$ SYM upon sending all auxiliary couplings to zero, we wish to carry out calculations using the smallest acceptable $\mu$ and $G$ / $\ka$.
If these are made too small, however, the lattice calculations will exhibit instabilities, such as the U(1) monopole condensation transition discussed in \refcite{Catterall:2014vka}.
The minimum acceptable values of the auxiliary couplings depend on both the lattice volume $L^4$ and the 't~Hooft coupling $\lalat$: We must increase $\mu$ and $G$ / \ka to consider larger $\lalat$, but can decrease them as $L$ increases.
In addition, for any given $(L, \lalat)$ the minimum acceptable value of $G$ or \ka depends on $\mu$, and vice versa, leaving us with a multi-dimensional parameter space to explore.

We proceed by roughly mapping out the parameter space using a small $4^4$ lattice volume, which tends to be more susceptible to instabilities than are lattices with larger $L$.
For each 't~Hooft coupling of interest, we scan over small ranges of $G$ for a few values of $\mu$, based around initial estimates extrapolated from smaller values of $\lalat$.
For the improved action we preferentially reduce $\mu$ even if this requires larger $G$, to minimize soft supersymmetry breaking.
For the unimproved action we fix $\ka = 0.5$ as discussed in \refcite{Catterall:2014vka}.
Finally, when moving from $L = 4$ to larger lattice volumes we currently reduce $\mu \sim 1 / L$ while keeping $G$ or \ka fixed.
This scaling for $\mu$ can be motivated by interpreting $\mu$ as a mass for the imaginary U(1) modes, and is less aggressive than the $\mu^2 \propto 1 / L^4$ proposed in \refcite{Catterall:2014vka} on the grounds that these modes are constant in space and time.
While we have found that we can also reduce $G$ on larger volumes, in the next subsection we will see that this does not seem to be necessary to rapidly approach the continuum limit where the U(1) sector (and thus the plaquette determinant deformation) decouples from the SU($N$) target theory.

Further details of the lattice phase diagrams for the improved and unimproved actions are not relevant to the present study.
Instead, in \tab{tab:ensembles} we simply list the lattice ensembles that produce the results presented in the next subsection, with couplings chosen through the procedure summarized above.
In this paper we consider mainly gauge group U(2); initial explorations for $N = 3$ and 4 suggest that the same values of $\mu$ and $G$ / \ka should be used for any $N$.

\begin{table}[htbp]
  \centering
  \renewcommand\arraystretch{1.2} 
  \begin{tabular}{cccc|cccc}
    \multicolumn{4}{c|}{Improved action}  & \multicolumn{4}{|c}{Unimproved action}  \\
    Volume  & \lalat  & $\mu$ & $G$       & Volume  & \lalat  & $\mu$ & \ka         \\
    \hline
    $4^4$   & 0.5     & 0.4   & 0.02      & $4^4$   & 0.5     & 0.5   & 0.5         \\
    $4^4$   & 1.0     & 0.4   & 0.05      & $4^4$   & 1.0     & 0.4   & 0.5         \\
    $4^4$   & 2.0     & 0.8   & 0.05      & $4^4$   & 2.0     & 0.5   & 0.5         \\
    $4^4$   & 3.0     & 0.8   & 0.05      & $4^4$   & 3.0     & 0.5   & 0.5         \\
    $4^4$   & 4.0     & 0.8   & 0.05      & $4^4$   & 4.0     & 0.8   & 0.5         \\
    $4^4$   & 5.0     & 0.8   & 0.10      & $4^4$   & 5.0     & 0.8   & 0.5         \\
    $4^4$   & 6.0     & 0.8   & 0.10      & $4^4$   & 6.0     & 1.0   & 0.5         \\
    $4^4$   & 7.0     & 1.0   & 0.15      &         &         &       &             \\
    $4^4$   & 8.0     & 1.0   & 0.25      &         &         &       &             \\
    $6^4$   & 1.0     & 0.3   & 0.05      & $6^4$   & 1.0     & 0.3   & 0.5         \\
    $6^4$   & 2.0     & 0.5   & 0.05      &         &         &       &             \\
    $6^4$   & 3.0     & 0.5   & 0.05      &         &         &       &             \\
    $6^4$   & 4.0     & 0.5   & 0.05      &         &         &       &             \\
    $8^4$   & 1.0     & 0.2   & 0.05      & $8^4$   & 1.0     & 0.2   & 0.5         \\
    $8^4$   & 2.0     & 0.4   & 0.05      &         &         &       &             \\
    $8^4$   & 3.0     & 0.4   & 0.05      &         &         &       &             \\
    $8^4$   & 4.0     & 0.4   & 0.05      &         &         &       &             \\
    $12^4$  & 1.0     & 0.15  & 0.05      & $12^4$  & 1.0     & 0.15  & 0.5         \\
    $16^4$  & 1.0     & 0.1   & 0.05      &         &         &       &             \\
  \end{tabular}
  \caption{\label{tab:ensembles}Auxiliary couplings used to obtain the results in figures~\protect\ref{fig:sB_compare}--\protect\ref{fig:susy_pieces}, chosen through the procedure summarized in \protect\secref{sec:phase}.  The gauge group is U(2) in all cases.}
\end{table}

\subsection{Ward identity violations and continuum extrapolations} 
\begin{figure}[htbp]
  \centering
  \includegraphics[height=\figheight]{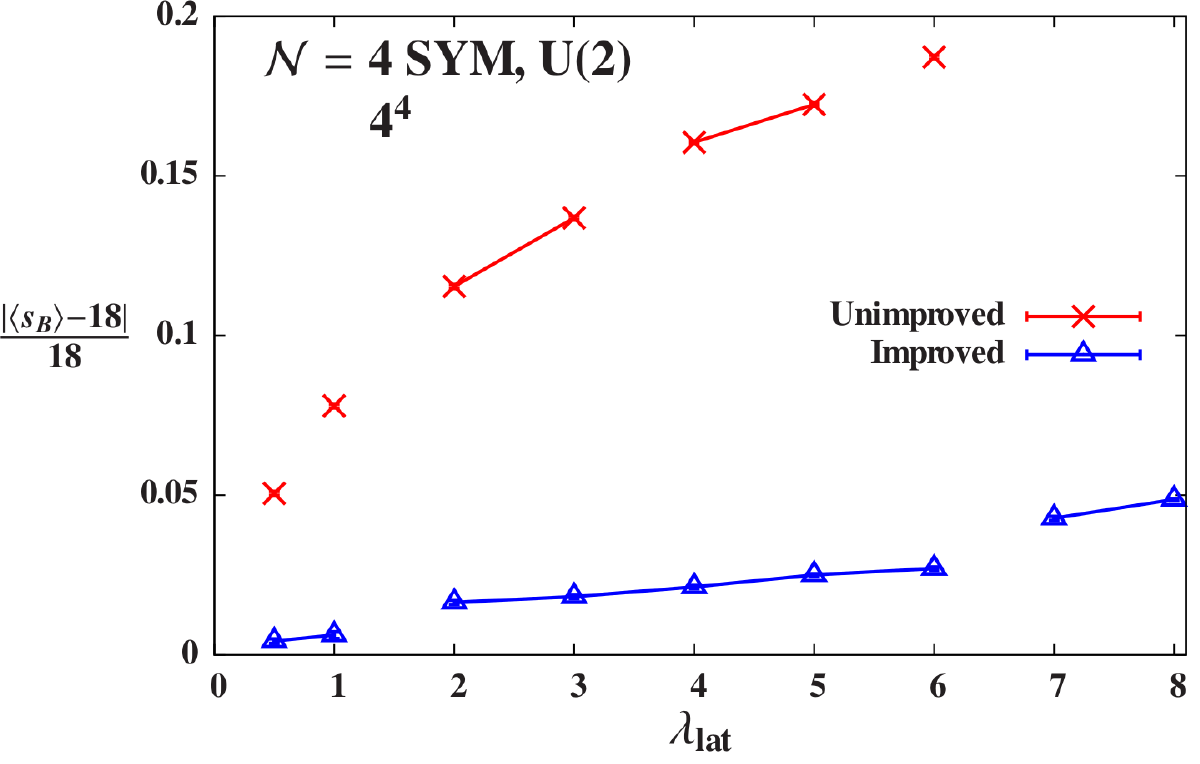}
  \caption{\label{fig:sB_compare}Deviations of the bosonic action from its exact supersymmetric value are much larger for $S_{\rm unimp}$ than for $S_{\rm imp}$ throughout a wide range of 't~Hooft coupling \lalat on $4^4$ lattices.  Lines connect points with fixed $\mu$ (cf.~\protect\tab{tab:ensembles}).}
\end{figure}

To quantify the severity of supersymmetry breaking, due to both soft breaking terms and the anti-periodic temporal BCs for the fermions, we will measure violations of \cQ Ward identities.
We begin with the Ward identity resulting from $\cQ S_{\rm formal} = 0$.
Because the fermion action is gaussian this Ward identity requires that the bosonic action per lattice site take the value $\vev{s_B} = 9N^2 / 2$ for gauge group U($N$), independent of the 't~Hooft coupling $\lalat$.
In \fig{fig:sB_compare} we plot normalized violations of this Ward identity for both $S_{\rm unimp}$ and $S_{\rm imp}$ across a wide range of \lalat on small $4^4$ lattices with $N = 2$.
Since we have to use larger $\mu$ as the coupling increases (\tab{tab:ensembles}) we can see that these violations are sensitive to both $\mu$ and $\lalat$, as expected.
The improved action produces roughly an order of magnitude reduction in the Ward identity violations, which allows us to reach significantly stronger couplings with percent-level \cQ breaking.

\begin{figure}[htbp]
  \centering
  \includegraphics[height=\figheight]{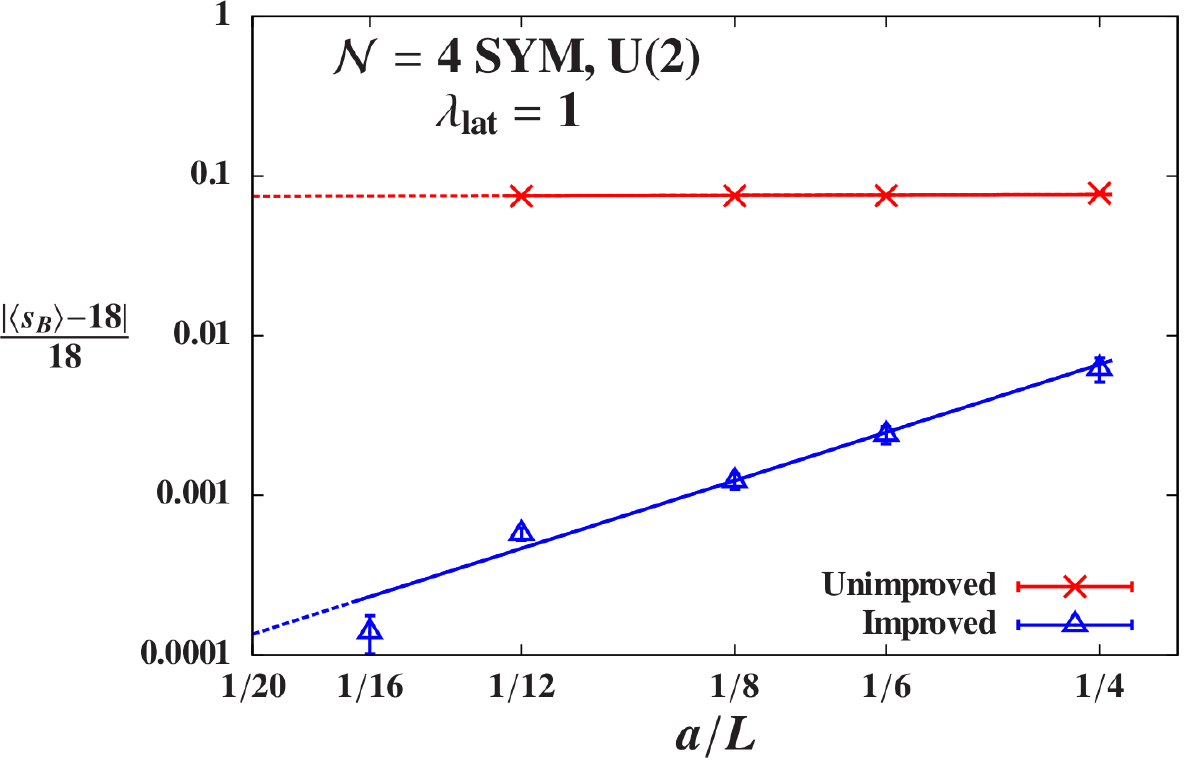}
  \caption{\label{fig:sB_power}Power-law continuum extrapolations of bosonic action deviations on logarithmic axes for $S_{\rm unimp}$ and $S_{\rm imp}$ with fixed $\lalat = 1$.  In both cases the deviations vanish $\propto (a / L)^p$ in the $(a / L) \to 0$ continuum limit, but the exponent $p = 2.42(13)$ for the improved action is much larger than the $p = 0.015(4)$ for the unimproved action.}
\end{figure}

\begin{figure}[htbp]
  \centering
  \includegraphics[height=\figheight]{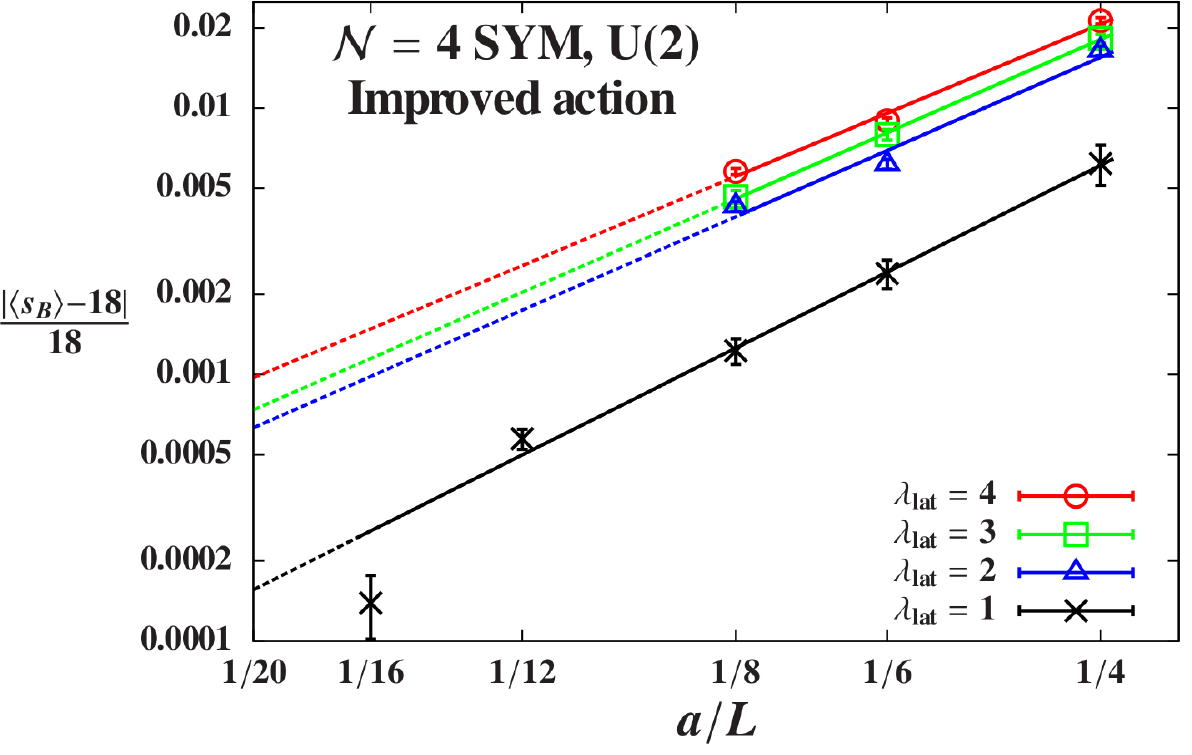}
  \caption{\label{fig:sB_power_lambda}Power-law continuum extrapolations of bosonic action deviations on logarithmic axes for the improved action with larger $2 \leq \lalat \leq 4$ in addition to the same $\lalat = 1$ points from \protect\fig{fig:sB_power}.  The deviations still vanish $\propto (a / L)^p$ at these stronger 't~Hooft couplings, with consistent $p \approx 2$ listed in \protect\tab{tab:exponents}.  The gap between $\lalat = 1$ and the other points is due to the larger values of $\mu$ that had to be used for $\lalat \geq 2$ (cf.~\protect\tab{tab:ensembles}).}
\end{figure}

The anti-periodic fermion BCs produce very little supersymmetry breaking for $S_{\rm imp}$, which was also the case for the unimproved action~\cite{Catterall:2014vka}.
However, the rest of the supersymmetry breaking from $\mu \neq 0$ is now so small that these BC effects can be relatively significant, at least on small $4^4$ volumes.
For example, switching from anti-periodic to periodic BCs for $(\lalat, \mu, G) = (1, 0.4, 0.1)$ roughly halves the normalized bosonic action deviations, from 0.0049(8) to 0.0021(5).

The improvement becomes even more striking when we consider larger lattice volumes, and in particular if we investigate the approach to the continuum limit as $(a / L) \to 0$.
In \fig{fig:sB_power} we plot the deviations of the bosonic action against $a / L$ on logarithmic axes, fixing $\lalat = 1$ for both $S_{\rm unimp}$ and $S_{\rm imp}$.
The right-most results on this plot match the $\lalat = 1$ points in \fig{fig:sB_compare}, and as $L$ increases the Ward identity violations decrease for both actions, as expected.
However the decrease is much more substantial for the improved action, which for $L = 16$ produces violations roughly 500$\times$ smaller than those from $S_{\rm unimp}$.

Fitting both sets of data in \fig{fig:sB_power} to power-law continuum extrapolations $\propto (a / L)^p$, we find $p = 2.42(13)$ for the improved action.
While the Ward identity violations from the unimproved action also vanish in the $(a / L) \to 0$ continuum limit, they approach this limit much more slowly, with $p = 0.015(4)$.
In \fig{fig:sB_power_lambda} we check that the improved action behaves consistently at stronger couplings, obtaining $p \approx 2$ as listed in \tab{tab:exponents}.
The $\sim$$(a / L)^2$ scaling of these results suggests that $S_{\rm imp}$ is effectively $\cO(a)$ improved, i.e.~the \cQ supersymmetry is preserved well enough to eliminate all non-negligible effects of dimension-5 operators.
In appendix~B we confirm that exact \cQ invariance, in combination with the other symmetries of lattice $\cN = 4$ SYM, would forbid all such dimension-5 operators.

\begin{figure}[htbp]
  \centering
  \includegraphics[height=\figheight]{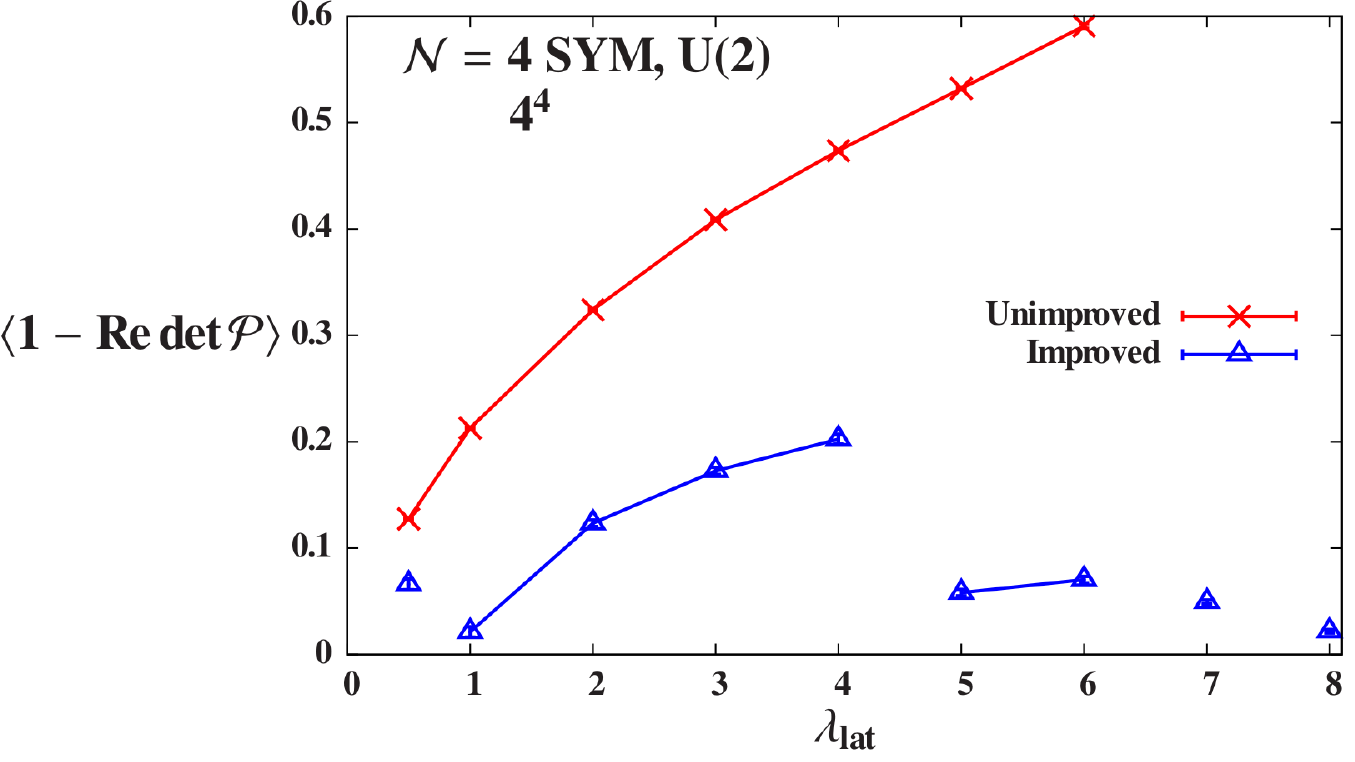}
  \caption{\label{fig:det_compare}$\vev{1 - \Re\det\cP} \to 0$ when the U(1) sector decouples from the SU($N$) target theory.  Here we average the plaquette determinant over all orientations and lattice sites.  This quantity is suppressed by one of the soft supersymmetry breaking terms in $S_{\rm unimp}$, while $S_{\rm imp}$ introduces $\vev{1 - \Re\det\cP} = 0$ as a new \cQ Ward identity when $G > 0$.  Across a wide range of \lalat on $4^4$ lattices, the improved action leads to smaller $\vev{1 - \Re\det\cP}$ than the unimproved action produces.  Lines connect points with fixed \ka or $G$ (cf.~\protect\tab{tab:ensembles}).  While larger values of \ka or $G$ trivially suppress $\vev{1 - \Re\det\cP}$, we want to use the smallest acceptable values to stay as close as possible to undeformed $\cN = 4$ SYM.}
\end{figure}

\begin{figure}[htbp]
  \centering
  \includegraphics[height=\figheight]{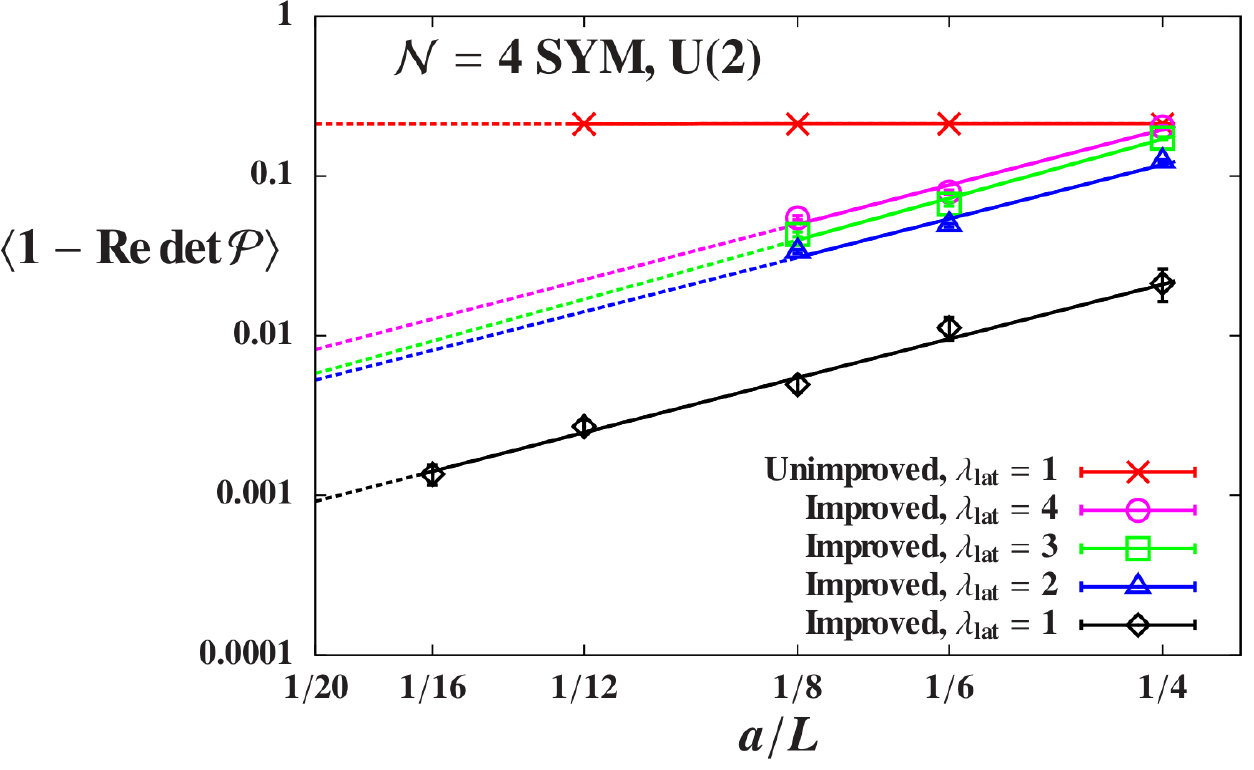}
  \caption{\label{fig:det_power}Power-law continuum extrapolations of $\vev{1 - \Re\det\cP}$ on logarithmic axes for the unimproved action with $\lalat = 1$ and the improved action with $1 \leq \lalat \leq 4$.  For $S_{\rm imp}$ this quantity vanishes $\sim$$(a / L)^2$ for each coupling (\protect\tab{tab:exponents}), as expected when the U(1) sector decouples from the SU($N$) target theory.  The gap between $\lalat = 1$ and the stronger couplings is due to the larger values of $\mu$ that had to be used for $\lalat \geq 2$ (cf.~\protect\tab{tab:ensembles}).  For $S_{\rm unimp}$ the corresponding exponent $p = 0.0027(21)$ is not significantly non-zero, implying that $L > 12$ is needed to reliably observe unimproved simulations approaching the SU($N$) continuum theory.}
\end{figure}

Next, as discussed in \secref{sec:deform}, the deformation of the auxiliary field equations of motion also introduces a new \cQ Ward identity for the improved action when $G$ is non-zero: $\vev{1 - \Re\det\cP} = 0$ from \eq{eq:detWard}.
Here we average the plaquette determinant over all orientations and lattice sites.
It is a gauge-invariant quantity associated with the U(1) sector, and approaches unity as this U(1) sector decouples from the SU($N$) target theory.
While the unimproved action does not protect the plaquette determinant through a Ward identity, one of the terms in \eq{eq:Ssoft} suppresses $\vev{1 - \Re\det\cP}$ at the cost of soft supersymmetry breaking.
In \fig{fig:det_compare} we plot $\vev{1 - \Re\det\cP}$ for $S_{\rm unimp}$ and $S_{\rm imp}$ across a wide range of \lalat on small $4^4$ lattices.
While the improved action produces smaller values for this quantity, on this lattice volume the results from the two actions are not dramatically different, especially at weaker couplings.
From \fig{fig:det_compare} and \tab{tab:ensembles} we can also see that larger values of the auxiliary coupling $G$ trivially reduce $\vev{1 - \Re\det\cP}$.
However, we still want to minimize $G$ to remain as close as possible to undeformed $\cN = 4$ SYM.

The real contrast between $S_{\rm unimp}$ and $S_{\rm imp}$ appears when we consider the $(a / L) \to 0$ continuum limit in \fig{fig:det_power}.
As for the bosonic action deviations, the improved action produces $\vev{1 - \Re\det\cP} \propto (a / L)^2$ for all investigated 't~Hooft couplings $1 \leq \lalat \leq 4$.
The specific exponents $p$ resulting from power-law fits $\propto (a / L)^p$ are collected in \tab{tab:exponents}.
For the unimproved action with $\lalat = 1$ it is not even clear that the U(1) sector is actually decoupling as $L$ increases within the range $4 \leq L \leq 12$.
The corresponding power-law fit produces an exponent $p = 0.0027(21)$ that is barely distinguishable from zero.

\begin{figure}[bp] 
  \centering
  \includegraphics[height=\figheight]{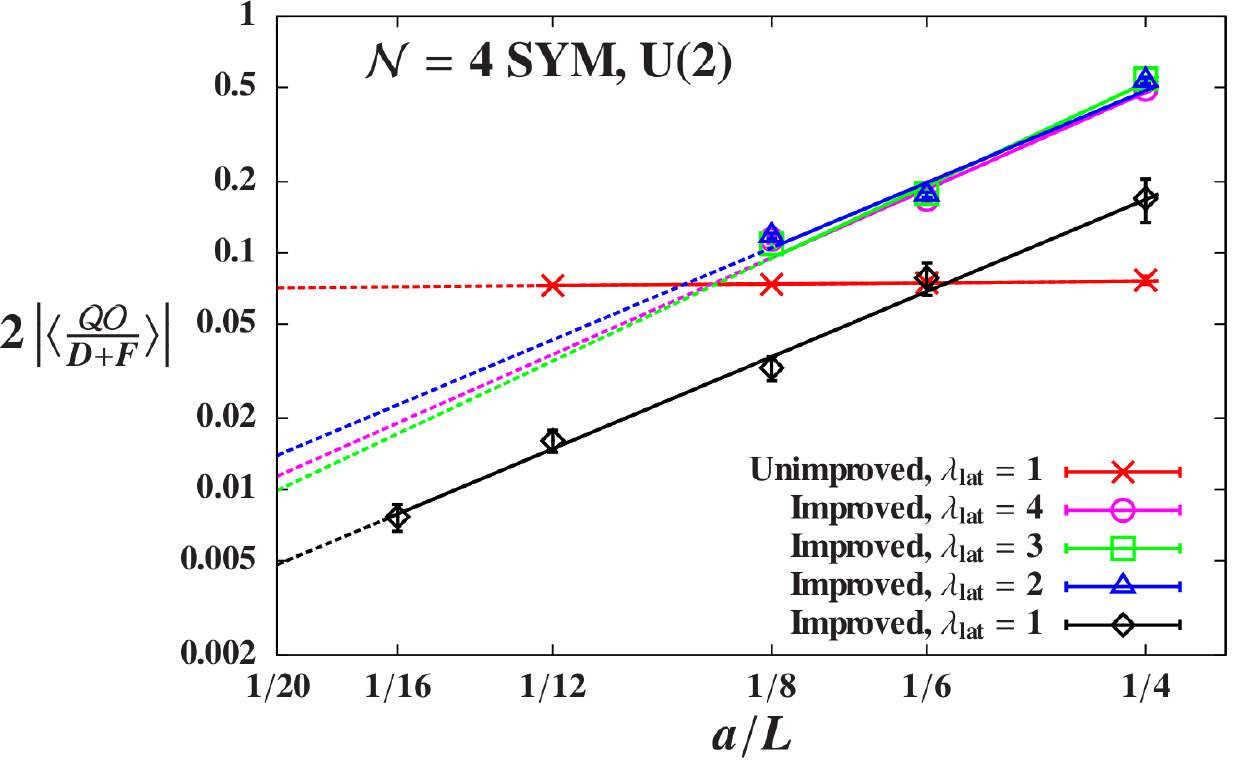}
  \caption{\label{fig:susy_power}Power-law continuum extrapolations of normalized $\eta\psi_a$ bilinear Ward identity violations (\protect\eq{eq:bilin}) on logarithmic axes for the unimproved action with $\lalat = 1$ and the improved action with $1 \leq \lalat \leq 4$.  This quantity vanishes $\propto (a / L)^p$ in the $(a / L) \to 0$ continuum limit, with $p = 0.040(17)$ for the unimproved action.  For the improved action larger violations on small lattice volumes are quickly compensated by the much larger $p \approx 2$ collected in \protect\tab{tab:exponents}.  The gap between $\lalat = 1$ and the stronger couplings is due to the larger values of $\mu$ that had to be used for $\lalat \geq 2$ (cf.~\protect\tab{tab:ensembles}).}
\end{figure}

\begin{table}[htbp]
  \centering
  \renewcommand\arraystretch{1.2}  
  \begin{tabular}{c|ccc}
                                & $\displaystyle \frac{|\vev{s_B} - 9N^2 / 2|}{9N^2 / 2}$ & $\vev{1 - \Re\det\cP}$  & $\displaystyle 2\left|\vev{\frac{\cQ \cO}{D + F}}\right|$ \\[8 pt]
    \hline
    ~Unimproved, $\lalat = 1$~  & 0.015(4)                                                & 0.0027(21)              & 0.040(17)                                                 \\
     Improved, $\lalat = 1$     & 2.42(13)                                                & 1.94(15)                & 2.21(15)                                                  \\
     Improved, $\lalat = 2$     & 1.99(9)                                                 & 1.93(6)                 & 2.20(7)                                                   \\
     Improved, $\lalat = 3$     & 1.99(9)                                                 & 2.10(6)                 & 2.46(7)                                                   \\
     Improved, $\lalat = 4$     & 1.90(6)                                                 & 1.97(6)                 & 2.32(6)                                                   \\
  \end{tabular}
  \caption{\label{tab:exponents}Results for the exponents $p$ obtained by fitting the listed lattice artifacts to power-law continuum extrapolations $\propto (a / L)^p$ as shown in figures~\protect\ref{fig:sB_power}, \protect\ref{fig:sB_power_lambda}, \protect\ref{fig:det_power} and \protect\ref{fig:susy_power}.  All exponents for the improved action are $p \approx 2$, suggesting that we have achieved $\cO(a)$ improvement as discussed in the text and in appendix~B.  Exponents for the unimproved action are much smaller, especially for $\vev{1 - \Re\det\cP}$, which is only protected by a \cQ Ward identity for $S_{\rm imp}$ with non-zero $G$.}
\end{table}

\begin{figure}[htbp]
  \centering
  \includegraphics[height=\figheight]{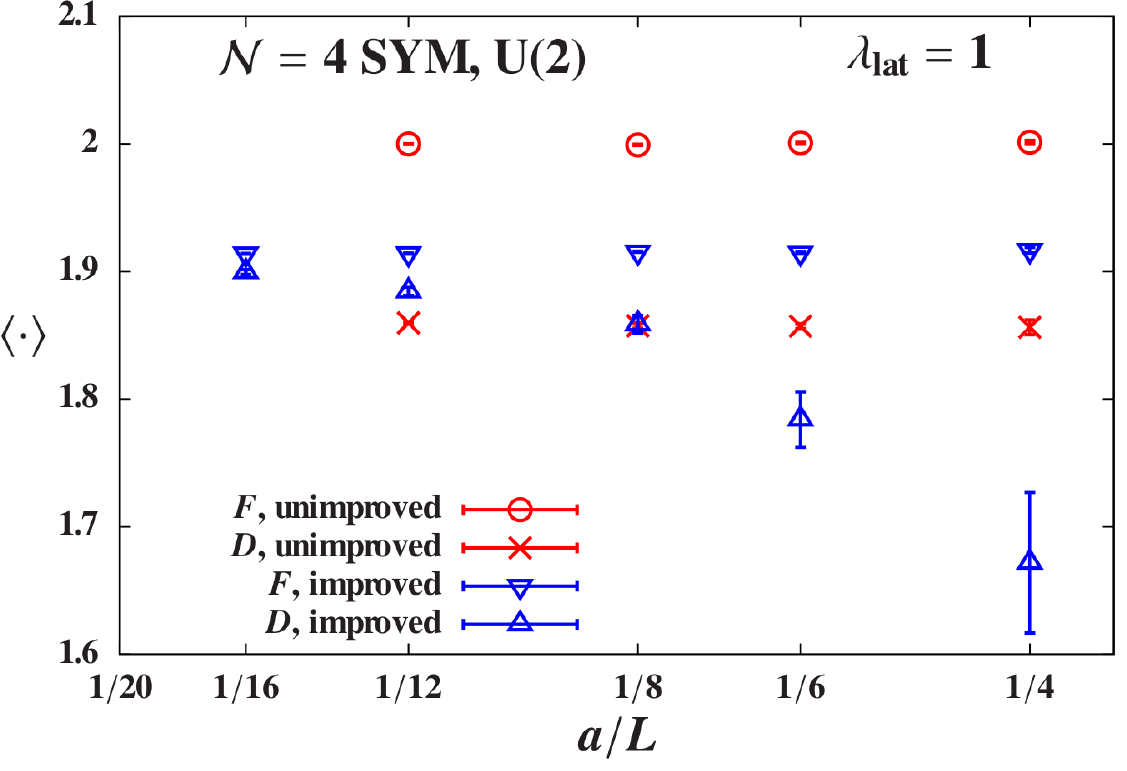}
  \caption{\label{fig:susy_pieces}Comparison of the $D$ and $F$ terms in \protect\eq{eq:bilin} for the $\eta\psi_a$ bilinear Ward identity at fixed $\lalat = 1$.  For the unimproved action both terms are fairly insensitive to the volume, with comparable uncertainties.  The fermion bilinear $F$ term behaves similarly for the improved action, while the pure-gauge $D$ term changes much more significantly and ends up with much larger jackknife uncertainties due to the modified equations of motion for the auxiliary field $d$ (\protect\eq{eq:EOM}).}
\end{figure}

The last Ward identity we consider involves the $\eta\psi_a$ fermion bilinear and was introduced in \refcite{Catterall:2014vka}: $\vev{\cQ \cO} = 0$ where $\cO = \Tr{\eta \sum_a \cU_a \cUbar_a}$ and so
\begin{equation}
  \label{eq:bilin}
  \cQ \cO = \Tr{d \sum_a \cU_a \cUbar_a} - \Tr{\eta \sum_a \psi_a \cUbar_a} \equiv D - F.
\end{equation}
We use the shorthand ``$D$'' and ``$F$'' because the first term depends on the modified equations of motion for the auxiliary field $d$, \eq{eq:EOM}, while the second involves the fermion bilinear.
In \fig{fig:susy_power} we show the usual power-law continuum extrapolations of the violations of this Ward identity, normalized by $\frac{1}{2}(D + F)$.
As for the other quantities considered above, the improved action produces Ward identity violations that vanish $\sim$$(a / L)^2$ in the continuum limit, while the exponent for the unimproved action is much smaller, $p = 0.040(17)$.
Precise values of $p$ for the improved action at each $1 \leq \lalat \leq 4$ are collected in \tab{tab:exponents}.

One curious feature of \fig{fig:susy_power} is that the improved action leads to much larger $\vev{\cQ \cO}$ on small lattice volumes, especially $4^4$.
Although this is quickly compensated by the more rapid approach to the continuum limit as $L$ increases, it is a striking change compared to figures~\ref{fig:sB_power} and \ref{fig:det_power}.
Contrary to our expectations, we find that the fermion bilinear term $F$ remains fairly insensitive to the volume for the improved action.
It is the pure-gauge term $D$ that now changes significantly as $L$ increases, steadily approaching $F$ as shown in \fig{fig:susy_pieces}.
The improved action also leads to large fluctuations in the $D$ term, producing jackknife uncertainties more than an order of magnitude larger than those of $F$ (even though we use only three stochastic sources to measure $F$ on each saved gauge configuration).
Not surprisingly given these results, the fermion BCs have no statistically significant effect on these Ward identity violations, which change from 0.056(2) to 0.063(9) when we switch from anti-periodic to periodic BCs for $(\lalat, \mu, G) = (1, 0.4, 0.1)$ on $4^4$ lattices.

\begin{table}[htbp]
  \centering
  \renewcommand\arraystretch{1.2}  
  \begin{tabular}{c|ccc}
            & $\displaystyle \frac{|\vev{s_B} - 9N^2 / 2|}{9N^2 / 2}$ & $\vev{1 - \Re\det\cP}$  & $\displaystyle 2\left|\vev{\frac{\cQ \cO}{D + F}}\right|$ \\[8 pt]
    \hline
    ~U(2)~  & 0.0062(11)                                              & 0.0212(49)              & 0.0849(176)                                               \\
     U(3)   & 0.0028(5)                                               & 0.0060(19)              & 0.0254(57)                                                \\
     U(4)   & 0.0024(2)                                               & 0.0036(10)              & 0.0162(31)                                                \\
  \end{tabular}
  \caption{\label{tab:susy_N}Bosonic action deviations, plaquette determinant fluctuations and $\eta\psi_a$ bilinear Ward identity violations from the improved action for gauge groups U($N$) with $N = 2$, 3 and 4 on $4^4$ lattices with fixed $(\lalat, \mu, G) = (1, 0.4, 0.05)$.  The results are reasonably consistent with $1 / N^2$ suppression; some deviations from such scaling are attributable to the anti-periodic fermion temporal BCs.}
\end{table}

Finally, in \refcite{Catterall:2014vka} we reported that Ward identity violations for the unimproved action decrease $\propto 1 / N^2$ for gauge group U($N$).
While all of the results discussed above are for U(2), our initial explorations of larger $N$ indicate that the improved action also exhibits this $1 / N^2$ suppression of supersymmetry breaking.
In \tab{tab:susy_N} we collect the bosonic action deviations, $\vev{1 - \Re\det\cP}$, and $\eta\psi_a$ bilinear Ward identity violations from improved $4^4$ calculations with $N = 2$, 3 and 4 at fixed $(\lalat, \mu, G) = (1, 0.4, 0.05)$.
These results are reasonably consistent with $1 / N^2$ scaling, though some of them become so small that they appear sensitive to the anti-periodic fermion temporal BCs.

\section{Conclusion} 
In this paper we have presented a procedure for lifting the U(1) flat directions in lattice $\cN = 4$ SYM in a way that preserves the exact lattice supersymmetry.
These U(1) flat directions are a necessary feature of the supersymmetric lattice construction, and must be regulated to ensure that the path integral is well defined and for the lattice theory to have the correct naive continuum limit.
Our procedure involves a deformation of the moduli space of the lattice theory, based on modifying the equations of motion of an auxiliary field.

We applied this procedure to construct a new lattice action that we compared against the ``unimproved'' action used in earlier work.
We saw that the new action reduces violations of supersymmetric Ward identities by up to an order of magnitude on small $L^4$ lattice volumes, and more importantly we found that these violations now fall rapidly towards zero $\propto 1 / L^2$ as $L$ increases.
Indeed, we argue that the new action is effectively $\cO(a)$-improved since there are no dimension-5 operators compatible with the lattice symmetries.
These numerical benefits appear to be well worth the roughly doubled computational expense, and we recommend that the improved action introduced in this work be used in future studies of lattice $\cN = 4$ SYM.

\section*{Acknowledgments} 
We thank Aarti Veernala, Joel Giedt, Poul Damgaard and Tom DeGrand for helpful conversations and continuing collaboration on lattice $\cN = 4$ SYM.
This work was supported by the U.S.~Department of Energy (DOE), Office of Science, Office of High Energy Physics, under Award Numbers DE-SC0008669 (DS) and DE-SC0009998 (SC, DS). 
Numerical calculations were carried out on the HEP-TH cluster at the University of Colorado and on the DOE-funded USQCD facilities at Fermilab.

\section*{Appendix A: Another possible lattice action and its difficulties} 
\addcontentsline{toc}{section}{Appendix A: Another possible lattice action and its difficulties}
\renewcommand{\thesection}{A}
As discussed at the end of \secref{sec:improved}, it is enticing to consider the possibility of constructing a lattice action that includes both of the necessary scalar potential and plaquette determinant deformations while still maintaining exact \cQ supersymmetry.
This requires making each term non-negative so that it cannot cancel out the other, leading us to consider
\begin{align}
  \cQ\; \Tr{\eta(n) \left(\cDbar_a^{(-)}\cU_a(n)\right)} \to \cQ\; \mbox{Tr}\Bigg[\eta(n) \Bigg(& \cDbar_a^{(-)}\cU_a(n) + G\sum_{a \neq b} \left|\det\cP_{ab}(n) - 1\right|^2 \Ibb_N        \\
                                                                                                & + B^2 \sum_a \left(\frac{1}{N} \Tr{\cU_a(n) \cUbar_a(n)} - 1\right)^2 \Ibb_N\Bigg)\Bigg], \nn
\end{align}
with two tunable parameters $B$ and $G$.
Although we use the same symbol $G$ for its coupling, the plaquette determinant term has changed compared to \eq{eq:linear}.

The modified equations of motion for the auxiliary field $d$ are now
\begin{equation}
  \label{eq:overC_EOM}
  d(n) = \cDbar_a^{(-)}\cU_a(n) + G\sum_{a \neq b} \left|\det\cP_{ab}(n) - 1\right|^2 \Ibb_N + B^2 \sum_a \left(\frac{1}{N} \Tr{\cU_a(n) \cUbar_a(n)} - 1\right)^2 \Ibb_N,
\end{equation}
which produce the \cQ Ward identity
\begin{equation}
  \label{eq:overC_Ward}
  G \vev{\sum_n \sum_{a \neq b} \left|\det\cP_{ab}(n) - 1\right|^2} + B^2 \vev{\sum_n \sum_a \left(\frac{1}{N} \Tr{\cU_a(n) \cUbar_a(n)} - 1\right)^2} = 0
\end{equation}
from \eq{eq:Ward}.
Every term in these sums is non-negative, so the only way for this Ward identity to be satisfied is for each term to vanish independently on each lattice site.
That is, even though the lattice action is now fully $\cQ$-invariant (with periodic BCs), reaching a supersymmetric vacuum requires simultaneously satisfying at least $15V$ non-trivial constraints, where $V$ is the lattice volume.
We will soon see that this is not feasible in practice, which leads us to name this possibility `the over-constrained action'.
Written in full, it is
\begin{align}
  S_{\rm over}                 & = S_{\rm exact}^{\prime\prime} + S_{\rm closed}                                                                                                                                                       \label{eq:Sover} \\
  S_{\rm exact}^{\prime\prime} & = -S_{\rm det}^{\prime} - S_{\rm pot}                                                                                                                                                                              \nn \\
                               & \quad + \frac{N}{2\lalat} \sum_n \mbox{Tr}\Bigg[-\cFbar_{ab}(n)\cF_{ab}(n) - \chi_{ab}(n) \cD_{[a}^{(+)}\psi_{b]}^{\ }(n) - \eta(n) \cDbar_a^{(-)}\psi_a(n)                                                        \nn \\
                               & \hspace{-30 pt} + \frac{1}{2}\left(\cDbar_a^{(-)}\cU_a(n) + G \sum_{a \neq b} \left|\det\cP_{ab}(n) - 1\right|^2 \Ibb_N + B^2 \sum_a \left(\frac{1}{N}\Tr{\cU_a(n) \cUbar_a(n)} - 1\right)^2 \Ibb_N\right)^2\Bigg] \nn \\
  S_{\rm det}^{\prime}         & = \frac{N}{2\lalat} 2G \sum_n \Tr{\eta(n)} \sum_{a \neq b} \det\cP_{ab}(n) \left[\det\cP_{ba}(n) - 1\right]                                                                                                        \nn \\[-10 pt]
                               & \hspace{6.4 cm} \times \Tr{\cU_b^{-1}(n) \psi_b(n) + \cU_a^{-1}(n + \muhat_b) \psi_a(n + \muhat_b)}                                                                                                                \nn \\[10 pt]
  S_{\rm pot}                  & = \frac{N}{2\lalat} \frac{2B^2}{N} \sum_n \Tr{\eta(n)} \sum_a \left(\frac{1}{N}\Tr{\cU_a(n) \cUbar_a(n)} - 1\right) \Tr{\psi_a(n) \cUbar_a(n)}                                                                     \nn
\end{align}
with $S_{\rm closed}$ still given by \eq{eq:Sclosed}.

\begin{table}[htbp]
  \centering
  \renewcommand\arraystretch{1.2}  
  \begin{tabular}{cccc}
    \multicolumn{4}{c}{Over-constrained action} \\
    Volume  & \lalat  & $B$   & $G$             \\
    \hline
    $4^4$   &  0.5    & 0.5   & 0.10            \\
    $4^4$   &  1.0    & 0.4   & 0.05            \\
    $4^4$   &  2.0    & 0.5   & 0.05            \\
    $4^4$   &  3.0    & 0.5   & 0.10            \\
    $4^4$   &  4.0    & 0.5   & 0.15            \\
    $4^4$   &  5.0    & 0.5   & 0.20            \\
    $4^4$   &  6.0    & 0.5   & 0.25            \\
    $4^4$   &  7.0    & 0.5   & 0.25            \\
    $4^4$   &  8.0    & 0.5   & 0.30            \\
    $4^4$   &  9.0    & 0.5   & 0.30            \\
    $4^4$   & 10.0    & 0.5   & 0.35            \\
    $6^4$   &  1.0    & 0.3   & 0.05            \\
    $8^4$   &  1.0    & 0.2   & 0.05            \\
    $12^4$  &  1.0    & 0.15  & 0.05            \\
  \end{tabular}
  \caption{\label{tab:overC_ensembles}Auxiliary couplings used to obtain the results for the over-constrained action in figures~\protect\ref{fig:sB_compare_all}--\protect\ref{fig:susy_pieces_all}.}
\end{table}

\begin{figure}[htbp]
  \centering
  \includegraphics[height=\figheight]{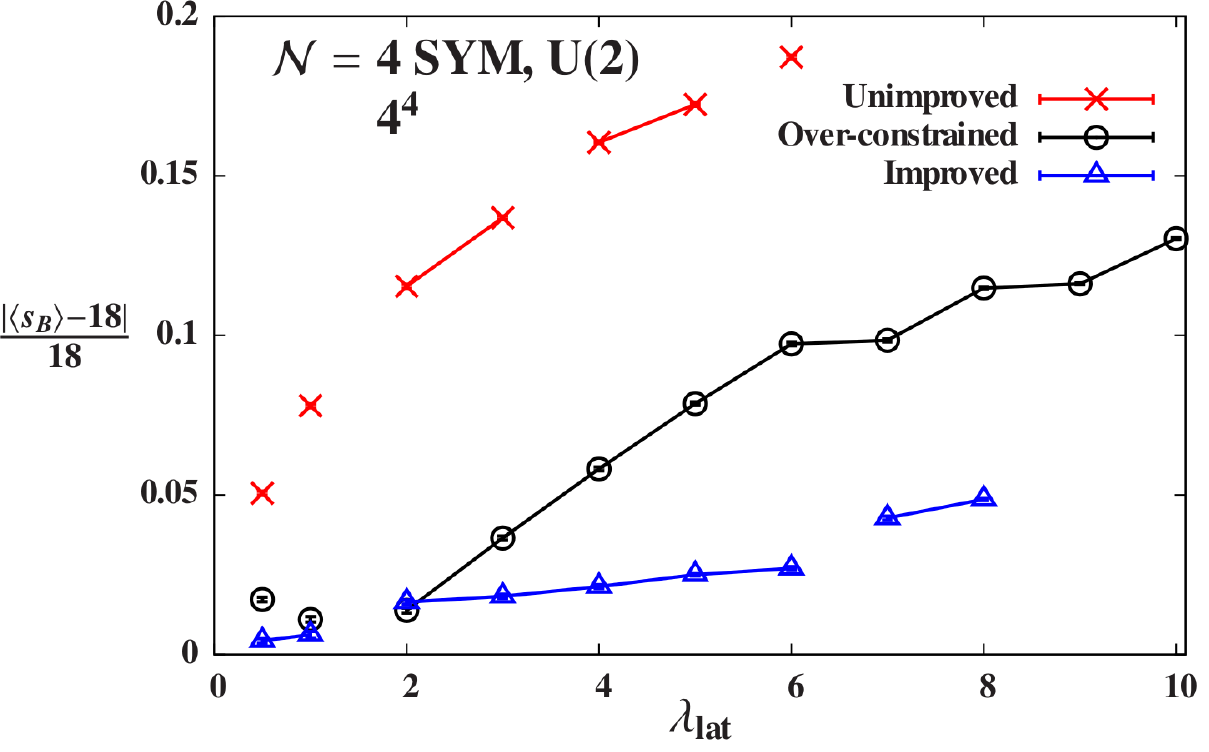}
  \caption{\label{fig:sB_compare_all}Deviations of the bosonic action from its exact supersymmetric value for the over-constrained action in addition to the other results in \protect\fig{fig:sB_compare} for a wide range of 't~Hooft coupling \lalat on $4^4$ lattices.  Although $S_{\rm over}$ leads to smaller deviations than the unimproved action produces, it does not behave as well as the improved action.  Lines connect points with fixed $B$ (cf.~\protect\tab{tab:overC_ensembles}).}
\end{figure}

\begin{figure}[htbp]
  \centering
  \includegraphics[height=\figheight]{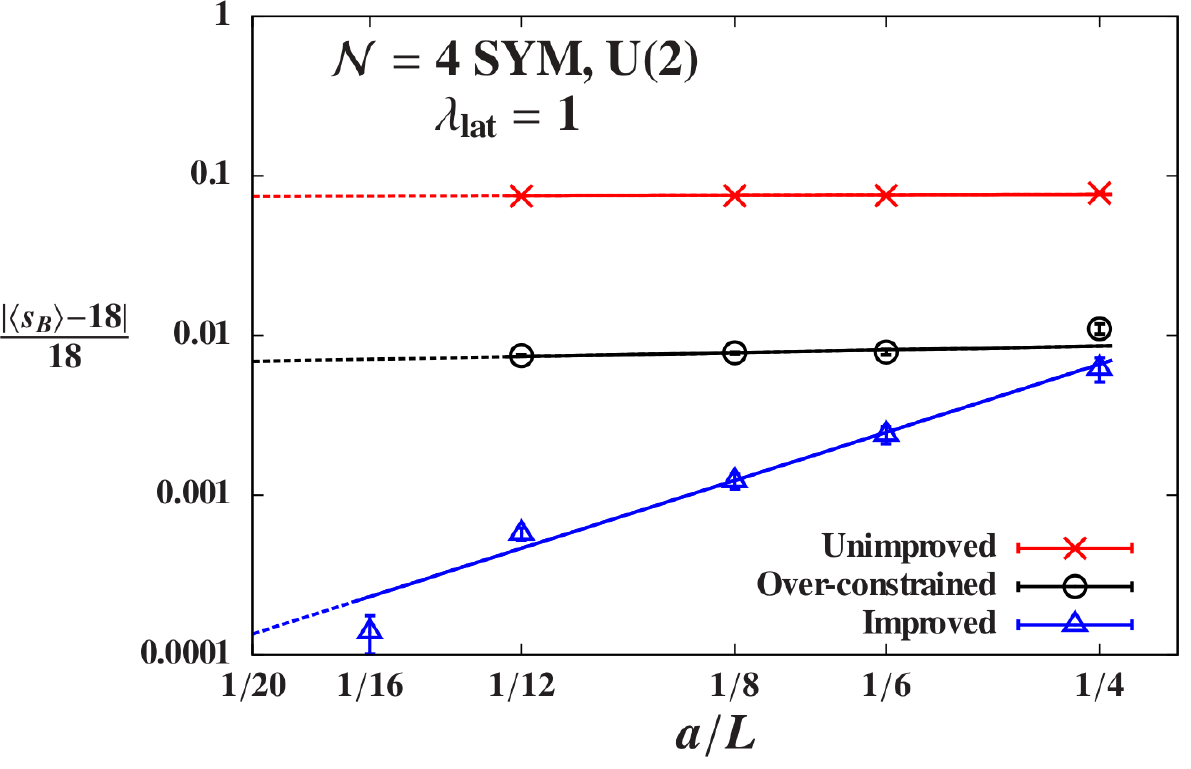}
  \caption{\label{fig:sB_power_all}Power-law continuum extrapolations of bosonic action deviations on logarithmic axes for the over-constrained action in addition to the other results in \protect\fig{fig:sB_power} at fixed $\lalat = 1$.  Although $S_{\rm over}$ produces $L = 4$ deviations comparable to those of the improved action, as $L$ increases these only decrease $\propto (a / L)^p$ with a much smaller power $p = 0.14(4)$.}
\end{figure}

\begin{figure}[htbp]
  \centering
  \includegraphics[height=\figheight]{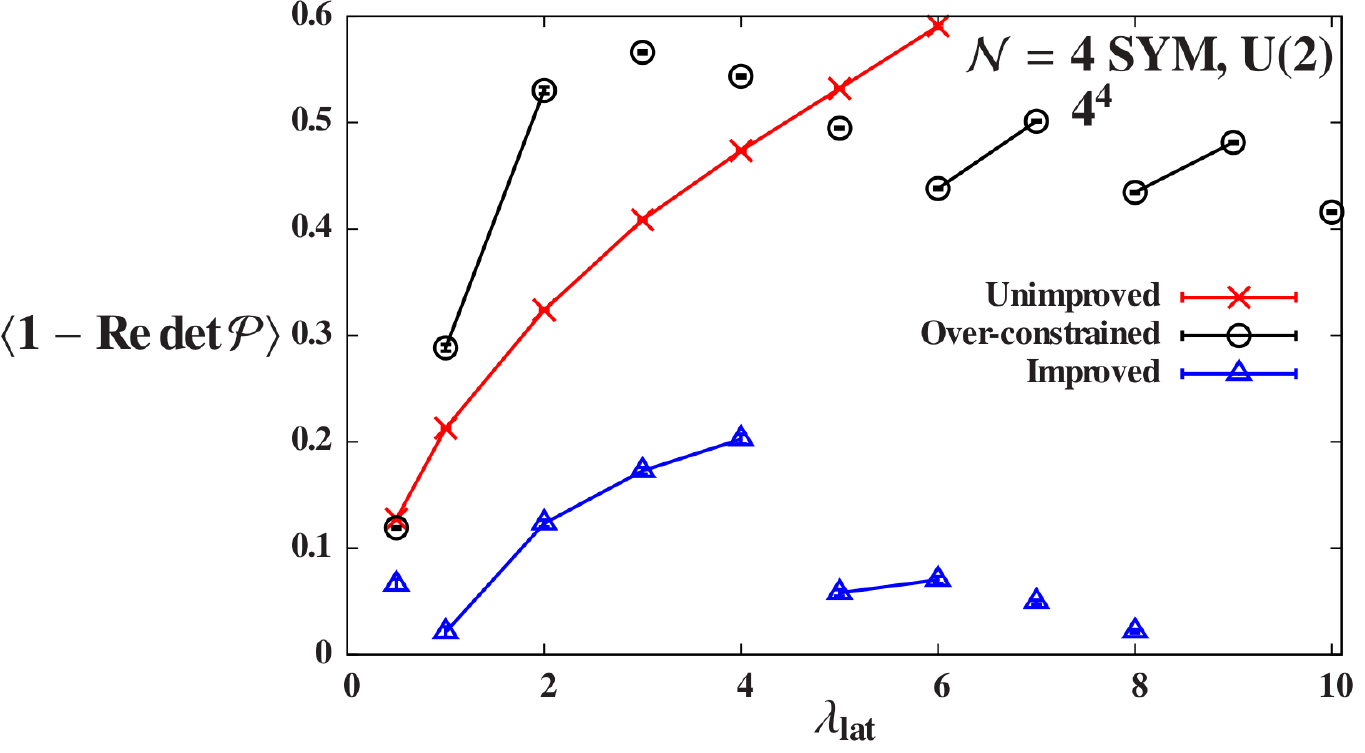}
  \caption{\label{fig:det_compare_all}$\vev{1 - \Re\det\cP}$ for the over-constrained action in addition to the other results in \protect\fig{fig:det_compare} for a wide range of 't~Hooft coupling \lalat on $4^4$ lattices.  Even though this quantity should be protected by a \cQ Ward identity (\protect\eq{eq:overC_Ward}), $S_{\rm over}$ often behaves worse than the unimproved action despite using relatively large $G$ (cf.~\protect\tab{tab:overC_ensembles}).  Lines connect points with fixed $G$.}
\end{figure}

\begin{figure}[htbp]
  \centering
  \includegraphics[height=\figheight]{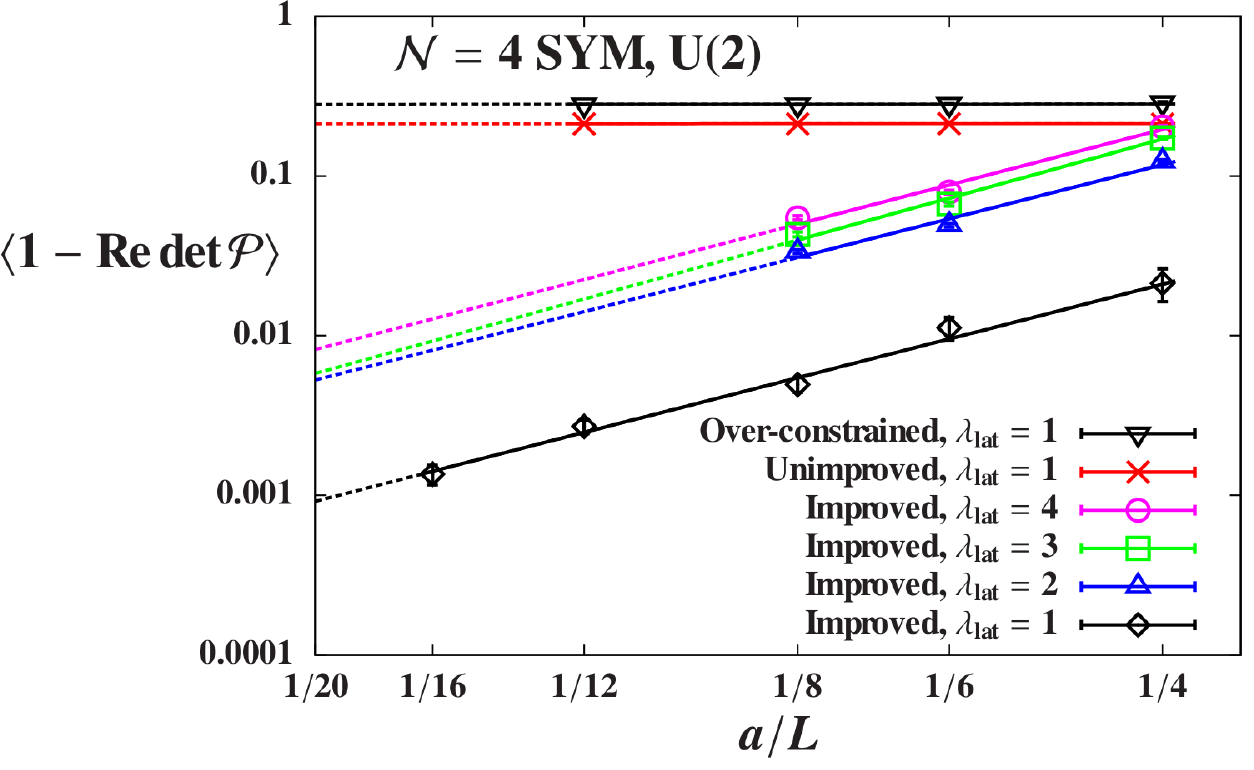}
  \caption{\label{fig:det_power_all}Power-law continuum extrapolations of $\vev{1 - \Re\det\cP}$ on logarithmic axes for the over-constrained action in addition to the other results in \protect\fig{fig:det_power}.  Not only does $S_{\rm over}$ produce larger plaquette determinant fluctuations even though this quantity should be protected by a \cQ Ward identity (\protect\eq{eq:overC_Ward}), the exponent $p = 0.0021(43)$ from fitting $\vev{1 - \Re\det\cP} \propto (a / L)^p$ is consistent with zero.}
\end{figure}

\begin{figure}[htbp]
  \centering
  \includegraphics[height=\figheight]{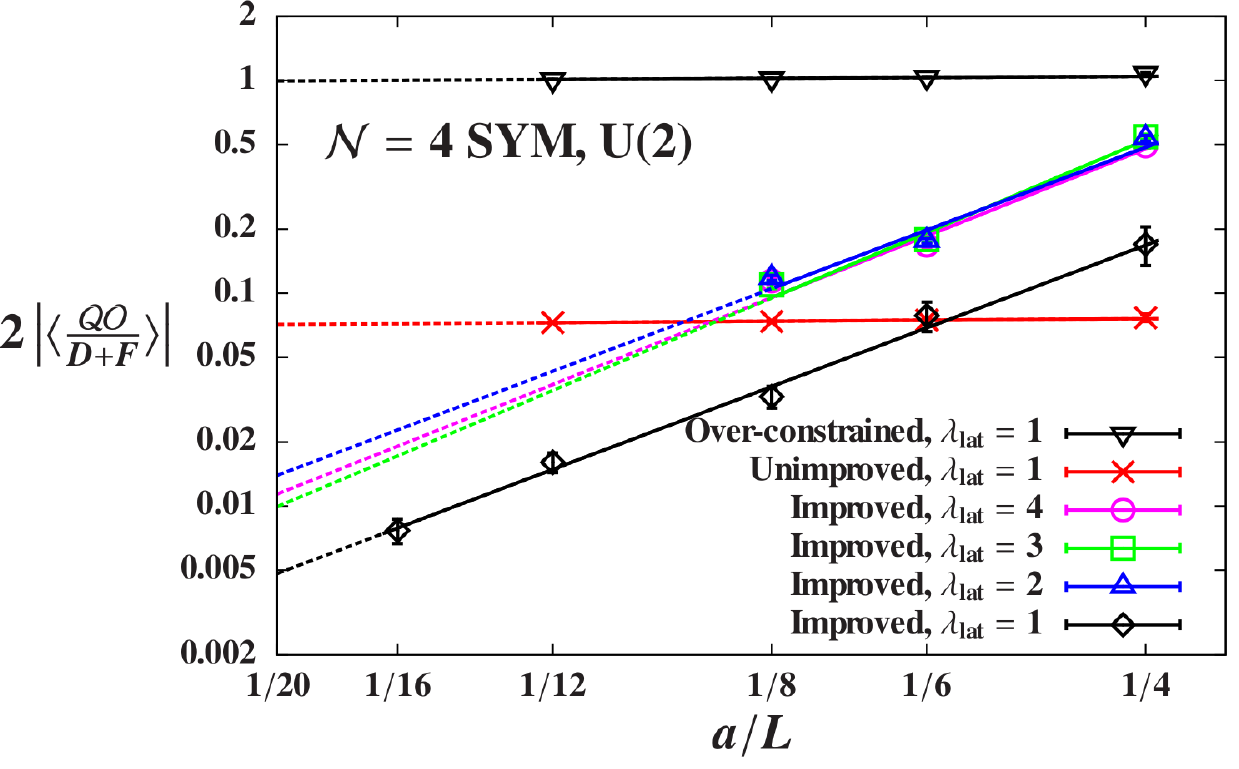}
  \caption{\label{fig:susy_power_all}Power-law continuum extrapolations of normalized $\eta\psi_a$ bilinear Ward identity violations (\protect\eq{eq:bilin}) on logarithmic axes for the over-constrained action in addition to the other results in \protect\fig{fig:susy_power}.  The modified equations of motion for $d$ in \protect\eq{eq:overC_EOM} produce very large $\cQ \cO = D - F$ that vanish very slowly in the continuum limit, $\propto (a / L)^p$ with $p = 0.031(1)$.}
\end{figure}

\begin{figure}[htbp]
  \centering
  \includegraphics[height=\figheight]{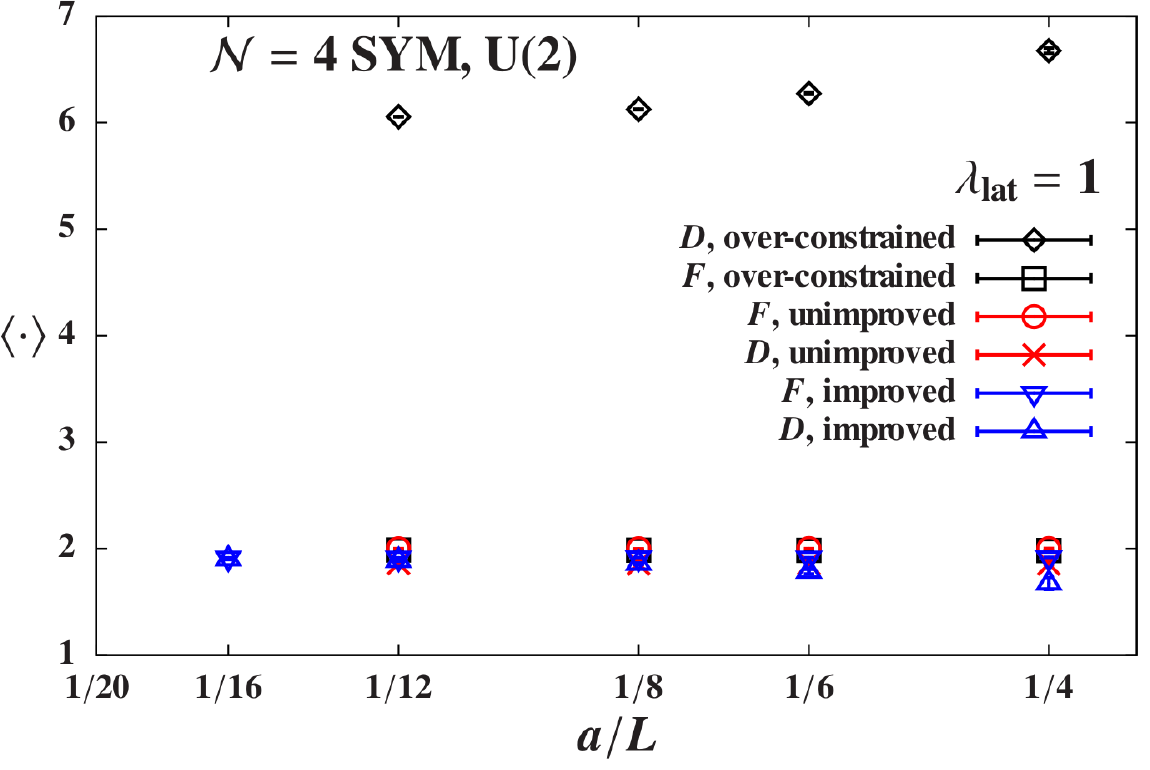}
  \caption{\label{fig:susy_pieces_all}Comparison of the $D$ and $F$ terms in the $\eta\psi_a$ bilinear Ward identity (\protect\eq{eq:bilin}) for the over-constrained action in addition to the other results in \protect\fig{fig:susy_pieces} at fixed $\lalat = 1$.  The modified equations of motion for $d$ in \protect\eq{eq:overC_EOM} produce pure-gauge terms $D$ that are now much larger than the fermion bilinear terms $F$, and only slowly decrease as $L$ increases.}
\end{figure}

With the over-constrained action defined, we can carry out the same parameter space exploration and investigation of the $(a / L) \to 0$ continuum limit as for the improved action in \secref{sec:tests}.
The resulting lattice ensembles are listed in \tab{tab:overC_ensembles} and produce the results shown in figures~\ref{fig:sB_compare_all}--\ref{fig:susy_pieces_all}.
Since these figures present the same quantities that we discussed at some length in \secref{sec:tests}, here we focus only on the new results from the over-constrained action.
First, the scan of bosonic action deviations across a wide range of 't~Hooft coupling \lalat on small $4^4$ lattices in \fig{fig:sB_compare_all} reveals that $S_{\rm over}$ still produces non-zero \cQ Ward identity violations despite avoiding all explicit supersymmetry breaking in the action.
In fact, $S_{\rm over}$ behaves worse than the improved action, though it still provides some benefit compared to $S_{\rm unimp}$.
Of course, the anti-periodic temporal BCs for the fermions do introduce some explicit \cQ breaking, but as for the other actions this is responsible for only a small portion of the deviations shown.
For $(\lalat, B, G) = (1, 0.5, 0.1)$ on $3^3\X 4$ lattices, switching from anti-periodic to periodic BCs only reduces the bosonic action deviation from 0.0260(11) to 0.0224(5).

The picture becomes worse in \fig{fig:sB_power_all}, where we consider the $(a / L) \to 0$ continuum extrapolation of the bosonic action deviations produced by the over-constrained action, and find that they vanish $\propto (a / L)^p$ with $p = 0.14(4)$.
While this exponent is an order of magnitude larger than that for the unimproved action (\tab{tab:exponents}), it is a far cry from the $p \approx 2$ obtained for the improved action.
That is, despite preserving the exact \cQ supersymmetry at the level of the action, $S_{\rm over}$ is not effectively $\cO(a)$ improved.

The situation is similar when we consider $\vev{1 - \Re\det\cP}$, again averaging the plaquette determinant over all orientations and lattice sites even though the Ward identity in \eq{eq:overC_Ward} should protect $\det\cP_{ab}(n) - 1$ site by site.
In \fig{fig:det_compare_all} we see that the over-constrained action often does a worse job controlling $\det \cP$ than the unimproved action, even though we end up having to use relatively large $G$ as listed in \tab{tab:overC_ensembles}.
We are also unable to observe the U(1) sector decoupling as we approach the continuum limit in \fig{fig:det_power_all}, where fitting $\vev{1 - \Re\det\cP} \propto (a / L)^p$ produces $p = 0.0021(43)$ indistinguishable from zero.
Since $\vev{1 - \Re\det\cP}$ should be protected by a \cQ Ward identity, this may also raise concerns about the restoration of the \cQ supersymmetry from $S_{\rm over}$ in the continuum limit.

Finally, we consider the $\eta\psi_a$ bilinear Ward identity violations in figures~\ref{fig:susy_power_all} and \ref{fig:susy_pieces_all}.
The over-constrained action clearly produces the worst results for this quantity, apparently due to the significant modifications of the auxiliary field equations of motion in \eq{eq:overC_EOM}.
These modifications lead to extremely large values for the pure-gauge term $D$ in \eq{eq:bilin}, while the fermion bilinear term $F$ from $S_{\rm over}$ is not changed so significantly.
Although $D$ does approach $F$ as the volume increases, the power-law fit $\propto (a / L)^p$ in \fig{fig:susy_power_all} still produces a very small $p = 0.031(1)$, worse even than the unimproved action.

In summary, despite maintaining exact \cQ supersymmetry the over-constrained action produces much more severe Ward identity violations than the improved action, and results in a much slower approach to the $(a / L) \to 0$ continuum limit.
The reason for this behavior seems to be that $S_{\rm over}$ requires the system to satisfy the $15V$ non-trivial constraints in \eq{eq:overC_Ward} in order to reach a supersymmetric vacuum, which is not possible in practice.
Since the Ward identities are violated even with fully periodic BCs, $S_{\rm over}$ appears to result in spontaneous supersymmetry breaking, which is forbidden in continuum $\cN = 4$ SYM where the Witten index is non-zero.
This suggests that the over-constrained lattice system is changing the Witten index by introducing new lattice states that don't exist in the continuum, which may be related to the slow approach to the continuum limit seen in figures~\ref{fig:sB_power_all}, \ref{fig:det_power_all} and \ref{fig:susy_power_all}.

While it may be worthwhile to continue searching for ways to further reduce or eliminate soft supersymmetry breaking in lattice $\cN = 4$ SYM, the alternate lattice action considered in this appendix is not viable.
In the context of practical numerical computations, the improved action introduced in \eq{eq:Simp} appears hard to beat.

\section*{Appendix B: Dimension-5 operators in lattice $\cN = 4$ SYM} 
\addcontentsline{toc}{section}{Appendix B: Dimension-5 operators in lattice $\cN = 4$ SYM}
\renewcommand{\thesection}{B}
\setcounter{equation}{0}
In this appendix we extend the enumeration of renormalizable operators in lattice $\cN = 4$ SYM from \refcite{Catterall:2014mha} to consider dimension-5 operators as well.\footnote{We thank Joel Giedt for discussions on this subject.}
In addition to respecting the \cQ supersymmetry, lattice gauge invariance and the $S_5$ point group symmetry of the $A_4^*$ lattice, these operators must also be uncharged under an additional U(1) ``ghost number'' symmetry~\cite{Catterall:2009it, Catterall:2014mha}.
This ghost number symmetry is hidden in the notation used to this point, and only becomes visible when the lattice fields are decomposed into irreducible representations of $S_5$ by means of the $5\X 5$ orthogonal matrix $P$ discussed in refs.~\cite{Unsal:2006qp, Catterall:2014vka}.
Schematically,
\begin{align}
  \cU_a   & \stackrel{P}{\to} (\cU_{\mu}, \phi)     & \cUbar_a  & \stackrel{P}{\to} (\cUbar_{\mu}, \phibar)         \\
  \psi_a  & \stackrel{P}{\to} (\psi_{\mu}, \etabar) & \chi_{ab} & \stackrel{P}{\to} (\chi_{\mu\nu}, \psibar_{\mu}) \nn
\end{align}
where $\mu = 0 \cdots 3$.
These fields have the ghost charges
\begin{table}[h]
  \centering
  \renewcommand\arraystretch{1.2} 
  \begin{tabular}{ccccccccc}
    ~$\cU_{\mu}$~ & ~$\phi$~  & ~$\cUbar_{\mu}$~  & ~\phibar~ & ~$\eta$~  & ~$\psi_{\mu}$~  & ~$\chi_{\mu\nu}$~ & ~$\psibar_{\mu}$~ & ~\etabar~ \\
     0            &  2        &  0                & -2        &  1        & -1              &  1                & -1                &  1.       \\
  \end{tabular}
\end{table}

Now we can begin by confirming that these symmetries permit no dimension-5 $\cQ$-closed operators.
Generic $\cQ$-closed terms involve only the fields that cannot be obtained through a supersymmetry transformation, namely $\eta$, $\cU_a$ and $\chi_{ab}$.
Two dimension-5 combinations of these fields are invariant under the $S_5$ point group symmetry:
\begin{align}
  & \chi_{ab} \cD_a^{(+)} \cD_b^{(+)} \eta &
  & \eps_{abcde}\ \cU_a \cU_b \cU_c \cU_d \cU_e.
\end{align}
The first of these terms vanishes because $\chi_{ab}$ is anti-symmetric under the interchange $a \leftrightarrow b$ while $\cD_a^{(+)} \cD_b^{(+)}$ is symmetric.
The \eps tensor in the other term requires that $\phi$ appear, leading to a non-zero ghost charge that forbids the operator.

Dimension-5 $\cQ$-exact operators can take the form $\cQ\; \Tr{\Psi_1 \Psi_2 \Psi_3}$ in addition to the $\cQ\; \Tr{\Psi f(\cU_a, \cUbar_a, d)}$ and $\cQ \left\{\Tr{\eta} \Tr{f(\cU_a, \cUbar_a, d)}\right\}$ considered in \refcite{Catterall:2014mha}.
Here each $\Psi$ stands for one of the fermion fields $\eta$, $\psi_a$ or $\chi_{ab}$ and $f(\cU_a, \cUbar_a, d)$ has to be a dimension-3 combination of the bosonic fields.
It is easy to see that there are no dimension-5 operators of the form $\cQ \left\{\Tr{\eta} \Tr{f(\cU_a, \cUbar_a, d)}\right\}$.
Such terms require an $S_5$-invariant $f(\cU_a, \cUbar_a, d) \to f(\cU_a \cUbar_a, d)$, which can only have $\dim[f] = 2$, 4 or higher.
While this also rules out $\cQ\; \Tr{\Psi f(\cU_a, \cUbar_a, d)}$ when $\Psi = \eta$, the $S_5$ symmetry allows terms with $\Psi = \psi_a$ and $\Psi = \chi_{ab}$:
\begin{align}
  & \cQ\; \Tr{\psi_a \cUbar_a d} &
  & \cQ\; \Tr{\psi_a \cUbar_a \cU_b \cUbar_b} &
  & \cQ\; \Tr{\eps_{abcde}\ \chi_{ab} \cUbar_c \cUbar_d \cUbar_e}.
\end{align}
The first term vanishes since \cQ annihilates each of $\psi_a$, $\cUbar_a$ and $d$ (\eq{eq:susy}), while the \eps tensor in the last term requires that \phibar appear, leading to a non-zero ghost charge.
The middle term also has a non-zero ghost charge, and was already discussed in \refcite{Catterall:2014mha}.
Finally, the only $S_5$-invariant $\cQ\; \Tr{\Psi_1 \Psi_2 \Psi_3}$ terms are
\begin{align}
  & \cQ\; \Tr{\eta\eta\eta} &
  & \cQ\; \Tr{\psi_a \psi_b \chi_{ab}},
\end{align}
both of which are forbidden by ghost charge conservation.

So we conclude that exact \cQ supersymmetry and the other symmetries of lattice $\cN = 4$ SYM forbid all dimension-5 operators.
When supersymmetry breaking is sufficiently small, we may therefore expect the system to be effectively $\cO(a)$ improved, explaining the behavior of the improved action observed in \secref{sec:tests}.

\section*{Appendix C: New results for the pfaffian phase} 
\addcontentsline{toc}{section}{Appendix C: New results for the pfaffian phase}
\renewcommand{\thesection}{C}
\setcounter{equation}{0}
The improved and over-constrained lattice actions both add the plaquette determinant to the fermion operator \cD in such a way as to break the $\eta \to \eta + c\Ibb_N$ shift symmetry and lift the corresponding U(1) fermion zero mode.
Cf.~$S_{\rm det}$ in \eq{eq:Simp} and $S_{\rm det}^{\prime}$ in \eq{eq:Sover}.
This has the serendipitous effect of allowing us to compute the pfaffian of \cD with fully periodic BCs, which is not possible for the formal or unimproved lattice actions.

As discussed in more detail by \refcite{Catterall:2014vka}, $\pf \cD = |\pf \cD| e^{i\al}$ results from integrating over the fermion fields in the lattice path integral and is potentially complex, $\al \neq 0$.
We carry out phase-quenched computations that determine observables $\vev{\cO}_{pq}$ without including the $e^{i\al}$ factor in the path integral, which we can then reintroduce through phase reweighting,
\begin{equation}
  \vev{\cO} = \frac{\vev{\cO e^{i\al}}_{pq}}{\vev{e^{i\al}}_{pq}}.
\end{equation}
In \refcite{Catterall:2014vka} we found that $e^{i\al} \approx 1$ with anti-periodic BCs, which makes the phase reweighting step irrelevant.
If the phase had fluctuated to the extent that $\vev{e^{i\al}}_{pq}$ became consistent with zero, then we would have faced a sign problem.

\begin{figure}[bp]
  \includegraphics[width=0.45\textwidth]{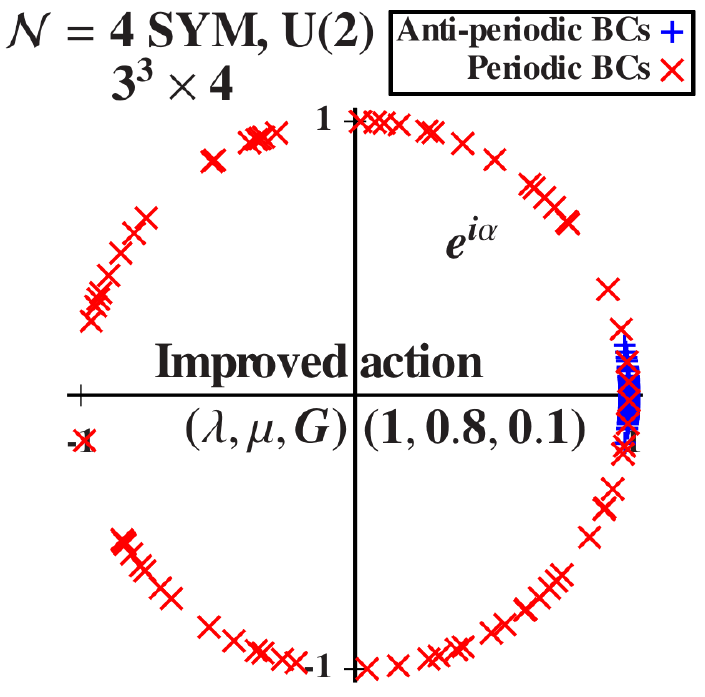}\hfill
  \includegraphics[width=0.45\textwidth]{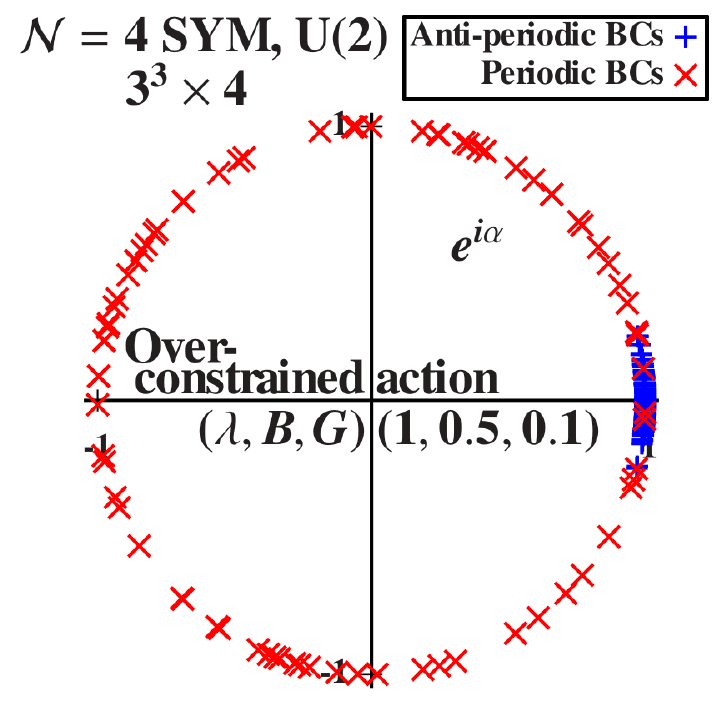}
  \caption{\label{fig:pfaffian}Scatter plots of pfaffian phase measurements from independent $3^3\X 4$ phase-quenched ensembles using either periodic or anti-periodic temporal BCs for the fermion fields, for the improved action with $(\lalat, \mu, G) = (1, 0.8, 0.1)$ (left) and the over-constrained action with $(\lalat, B, G) = (1, 0.5, 0.1)$ (right).  For both actions anti-periodic BCs produce $e^{i\al} \approx 1$ as we found for the unimproved action in \protect\refcite{Catterall:2014vka}.  Periodic BCs, however, lead to uncontrolled fluctuations and $\vev{e^{i\al}}_{pq}$ consistent with zero, indicating a sign problem.}
\end{figure}

Because we used the unimproved action in \refcite{Catterall:2014vka}, we were not able to investigate the pfaffian phase with periodic BCs.
In \fig{fig:pfaffian} we compare the pfaffian phase with periodic vs.\ anti-periodic BCs for both the improved action and the over-constrained action.
In both cases anti-periodic BCs produce $e^{i\al} \approx 1$ as we found for the unimproved action.
Taking the ensemble average,
\begin{align*}
  \vev{e^{i\al}}_{pq} & = 0.9977(4) + i0.0003(82)  \mbox{ for } S_{\rm imp} \\
  \vev{e^{i\al}}_{pq} & = 0.9952(9) + i0.0020(118) \mbox{ for } S_{\rm over}.
\end{align*}
With periodic BCs, however, the pfaffian phase fluctuates enough to average to zero for both actions, indicating a sign problem:
\begin{align*}
  \vev{e^{i\al}}_{pq} & = 0.071(82) - i0.029(88)  \mbox{ for } S_{\rm imp} \\
  \vev{e^{i\al}}_{pq} & = -0.041(80) + i0.086(78) \mbox{ for } S_{\rm over}.
\end{align*}

It is not yet clear to us why the pfaffian is so sensitive to the temporal fermion BCs.
Even more mysteriously, all other observables change very little between the ensembles generated with periodic or anti-periodic BCs, despite the apparent presence of a sign problem in the former case and its absence in the latter.
For example, the bosonic action deviations from the $S_{\rm imp}$ ensembles that produce the left plot in \fig{fig:pfaffian} are 0.0169(8) with anti-periodic BCs and 0.0115(8) with periodic BCs, just what we would expect in the absence of a sign problem.
This numerical evidence suggests that lattice $\cN = 4$ SYM does not suffer from a sign problem even when $\vev{e^{i\al}}_{pq}$ is consistent with zero.
We continue to search for ways to analytically understand this unexpected result.

\bibliographystyle{utphys}
\bibliography{improvedN4}
\end{document}